\documentclass[useAMS,usenatbib,usegraphicx]{mn2e}

\newcommand{\kms}{~km s$^{-1}$~}
\newcommand{\XSPEC}{{\small XSPEC~}}

\title[X-Rays From Massive OB Stars]{X-Rays From Massive OB Stars:
Thermal Emission From Radiative Shocks}

\author[S. A. Zhekov and F. Palla]
{Svetozar A. Zhekov$^{1}$\thanks{E-mail: szhekov@space.bas.bg} 
and Francesco Palla$^{2}$\thanks{E-mail: palla@arcetri.astro.it}\\
$^{1}$ Space Research Institute, Sofia-1000, Moskovska str. 6, Bulgaria\\
$^{2}$ INAF-Osservatorio Astrofisico di Arcetri, I-50125 Firenze, Italy}
\begin{document}

\date{}

\maketitle

\label{firstpage}

\begin{abstract}
{\it Chandra} gratings spectra of a sample of 15 massive OB stars were
analyzed under the basic assumption that the X-ray emission is produced 
in an ensemble of shocks formed in the winds driven by these objects. 
Shocks develop either as a result of radiation-driven instabilities 
or due to confinement of the wind by relatively strong magnetic field,
and since they are radiative, a simple model of their X-ray emission 
was developed that allows a direct comparison with observations.
According to our model, the shock structures (clumps, complete 
or fractional shells) eventually become `cold' clouds in the X-ray sky 
of the star. As a result, it is expected that for large covering factors
of the hot clumps, there is a high probability for X-ray
absorption by the `cold' clouds, resulting in blue-shifted
spectral lines. Our analysis has revealed that such a
correlation indeed exists for the considered sample of OB stars. 
As to the temperature
characteristics of the X-ray emission plasma, the studied OB stars
fall in two groups: (i) one with plasma temperature limited to
$\sim$0.1-0.4 keV; (ii) the other wtih X-rays 
produced in plasmas at considerably higher temperatures.
We argue that the two groups correspond to different mechanisms for 
the origin of X-rays: in
radiative-driven instability shocks and in
magnetically-confined wind shocks, respectively.
\end{abstract}


\section{Introduction} 
\label{sec:intro} 
The present concept on the origin of X-rays in massive OB stars
posits that they are emitted by hot gas heated by shocks 
(\citealt{lw_80}; \citealt{lu_82}). OB
stars possess massive and fast winds driven by radiation pressure
(e.g. \citealt{cak_75}) and subject to instabilities
(radiation-driven instabilities, RDI, \citealt{ow_84}) which
may give rise to the formation of strong shocks (\citealt{ow_88};
\citealt{feld_97}). The resulting
emission is expected to be relatively soft (plasma temperatures 
$kT < 1$~keV), while the
X-ray lines formed in the optically thick outflowing gas are predicted to
be blue-shifted and asymmetric due to the different absorption of the
red and blue part of the line profiles (\citealt{ig_01}; \citealt{ow_01}).

Recently, it has been suggested that the presence of magnetic
fields may be an important ingredient for the X-ray production
mechanism in very young massive stars (\citealt{ba_97}; \citealt{ud_02}).
The main physical effect of an ordered
magnetic field is to channel the stellar
wind towards the equatorial plane where the flows from the opposite stellar
hemispheres
collide, leading to the formation of strong shocks. The basic
feature of the so-called magnetically-confined wind shock model (MCWS) is that
the X-ray emission should be much harder (plasma temperatures 
$kT > 1$~keV) than that originating from the RDI shocks, and that
the X-ray line profiles should be narrower (line widths as small as
$\frac{1}{4}$ of the stellar wind velocity).

With the launch of the {\it Chandra} and {\it XMM-Newton}
observatories, the quality of the X-ray data on OB stars has improved
considerably, making it possible to test models of the origin of X-ray
emission.
Amongst the more than a dozen OB stars with gratings spectra of 
acceptable quality, 
only one object, $\zeta$~Puppis, 
shows unambiguous blueshifted spectral lines with 
asymmetric profiles (\citealt{cass_01}; \citealt{kahn_01}).
On the other hand,  there are two objects, $\theta^1$~Ori C 
(Schulz et al. 2000, 2003; \citealt{ga_05}) and 
$\tau$~Sco \citep{co_03}, in which a clear manifestation of the 
MCWS effects
is observed. While the scarcity of the MCWS-type
objects might be accounted for by their youth and the expected decay of
the magnetic field strength with the age of the massive star
\citep{schu_03}, the standard RDI shock model needs further
refinements to bring theory in accord with observations.

The fact that most OB stars show no appreciable line shifts and
asymmetric profiles is an obvious indication that their 
winds are much more transparent in X-rays than originally believed.
This result, along with
the small filling factors of the X-ray emitting plasma in OB stars as
found in previous ({\it ROSAT}) studies \citep{ku_96}, 
lends support to a physical picture in which the stellar winds 
are clumpy and/or porous. Alternatively, if the winds are smooth
and homogeneous, the mass-loss rates should have
values appreciably smaller (a factor $\sim$5 or more) 
than those presently accepted (\citealt{kr_03}; \citealt{co_06}). 

Wind clumping and porosity effects on the X-ray emission from OB stars,
and specifically on the shape of the line profiles, have been explored in 
recent analytical and numerical models 
(e.g., \citealt{feld_03}; Oskinova, Feldmeier \& Hamann 2004, 2006; 
\citealt{ow_06}).
It has been shown that
under given conditions (such as reduced mass loss and assumed `typical'
distance between clumps) their inclusion may lead to results much more 
consistent with the observations.
Given the complexity of these models and the fact that they are not yet fully
self-consistent, it is important to gather empirical information about 
the physical conditions in the
regions responsible for the X-ray emission to put further 
constraints on numerical models.

Various diagnostics have been used in this respect: analysis of helium-like
triplet ratios, global fits with discrete-temperature models,
constructing a distribution of emission measures as function of
temperature, based on fits to individual-line fluxes and to the total
X-ray spectra (\citealt{cass_01}; \citealt{kahn_01}; \citealt{wa_00};
\citealt{mi_02}; \citealt{co_03}; \citealt{schu_03}; \citealt{sa_04};
Wojdowski \& Schulz 2004, 2005; \citealt{ga_05}; \citealt{leu_06}). 
These studies
indicate that X-rays are produced close to the stellar surface (likely,
in the wind acceleration zone), and that the corresponding hot plasma has a
temperature stratification. The latter point reinforces the idea that the X-ray
emission originates in an ensemble of shocks.

Guided by this background, we have developed a simple model which bears all
the basic charateristics of the X-ray production in wind shocks.
The model is described 
in \S~\ref{sec:mod}; the data sample is given in \S~\ref{sec:obs}; 
the results are presented in \S~\ref{sec:res} and discussed in 
\S~\ref{sec:dis}. The conclusions close the paper.

\section{Model}
\label{sec:mod}

As in the case of the RDI and MCWS models, our basic assumption is that 
the X-ray emission of hot
massive stars originates in shocks. Given the relatively
high densities in the wind, the energy
losses by the shock-heated plasma are considerable: thus, shocks
should be {\it radiative}. 
This conclusion follows from simple estimates which show 
that the typical cooling time of a parcel of gas at the postshock 
temperature and density is smaller than the typical 
dynamic time of the 
flow. Namely, for the shock position at a distance $r$ from the star,
the ratio of the cooling time ($t_c = \frac{5}{2}p_{sh}/Q_c$)
 to the dynamic time of the flow ($t_d = r/v$) is: 
\begin{equation}
   \frac{t_c}{t_d} = 1.58\times10^{-2} T_{sh}^{1.6} 
     \left(\frac{r}{R_{\odot}}\right) v_{1000}^2 \dot{M}_6^{-1}
\label{eq:tcool}
\end{equation}
and the ratio of the thickness of the radiative shock (i.e.,
the cooling length of a parcel of hot gas at the postshock temperature:
$l_c = \frac{1}{4} v_{sh} t_c$) and shock `radius' is:
\begin{equation}
   \frac{l_c}{r} = 3.59\times10^{-3} T_{sh}^{2.1}
     \left(\frac{r}{R_{\odot}}\right) v_{1000} \dot{M}_6^{-1}
\label{eq:lcool}
\end{equation}
where $T_{sh}$ is the postshock temperature given in keV, 
$v_{1000}$ is the terminal stellar wind velocity (in units of 1000\kms),
and $\dot{M}_6$ is the stellar mass loss (in units of
$10^{-6}$~M$_\odot$~yr$^{-1}$). 
For a strong shock, the relation between the postshock temperature 
(in keV) and the
shock velocity (in units of 1000\kms) is given by 
$T_{sh} = 1.956\mu v_{shock}^2$, and $\mu$ is the mean atomic weight.
The relative number density of
hydrogen is assumed $x_H = 0.9$; thus, the relative electron
number density is $x_e = 1.1$ for a fully ionized plasma.
The gasdynamical quantities and the cooling function are described
in \S~\ref{subsec:shock}.

For typical mass-loss rates ($\dot{M}_6 \approx 0.1 - 1$) and  
wind velocities ($v_{1000} \approx 1.5 - 2.5$), the cooling time and 
cooling length of a shock developed in a wind are 
appreciably smaller than the corresponding typical characteristics of 
the flow (see eqs.~[\ref{eq:tcool}],~[\ref{eq:lcool}]).  This is true for 
postshock temperatures below 1 keV 
(i.e., shock velocities smaller than 1000\kms for solar abundances)
and shock locations from a few to several tens of stellar radii.
Therefore, the assumption of a steady-state, plane-parallel, radiative shock 
is a good approximation for our analysis. 
This conclusion finds support in 
numerical simulations of both the RDI and MCWS models.
In fact, as a result of 
efficient cooling, the shocks `collapse' in 
geometrically thin shells and disk-like structures, respectively
(e.g. \citealt{feld_97}; \citealt{ud_02}; \citealt{ga_05}).

The description of
our shock model and the ensemble of shocks is given below. 
The models were then used in the recent version (11.3.2) 
of the software package for analysis of X-ray spectra, \XSPEC 
\citep{a_96}.
A global-fit approach was adopted in our analysis for the following reasons:
(i) the fit can automatically take into account the
quasi-continuum due to numerous weak lines; (ii) by fitting the shape
of the underlying continuum, the model places additional constraints on 
the plasma temperature; (iii) the model can constrain the column density 
of the X-ray absorbing gas; and 
(iv) it can yield estimates of relative element abundances.
Finally, the X-ray emission is assumed to originate from a hot
optically-thin plasma in collisional ionization equilibrium (CIE),
as is the case for various types of astrophysical objects, including
hot massive stars (e.g., \citealt{pa_03}).

\begin{figure}
\includegraphics[width=3in, height=2.5in]{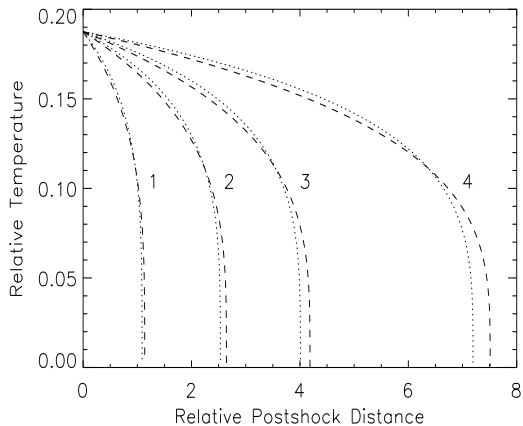}
\caption{Temperature profile in a steady-state radiative shock.
Dotted lines denote the exact solution, while the dashed ones give
the approximate solution under constant-pressure assumption. The four curves
are for shock velocities of 500, 650, 750 and 900\kms, 
as correspondingly labeled: 1, 2, 3, and 4.
}
\label{fig:rshock} 
\end{figure}

\subsection{Shock Model} 
\label{subsec:shock}
We consider a steady-state, plane-parallel, {\it radiative} shock moving
in a gas  with adiabatic index $\gamma = 5/3$. The basic 
physical quantities of the flow (density or nucleon number density, 
velocity, 
and pressure) have their standard postshock values for a strong shock: 
$\rho_{sh} = 4\rho_0$ ($n_{sh} = 4n_0$), $v_{sh} = v_0/4$, 
and $p_{sh} = 3/4\rho_0v_0^2$,
where the subscript `0' denotes the preshock values
(the gas velocity is given in the rest frame of the shock front).
The mass and momentum conservation equations
that describe the downstream flow are:
$\rho v = \rho_{sh} v_{sh} = \rho_0 v_0$ 
($n v = n_{sh} v_{sh} = n_0 v_0$),
$\rho v^2 + p = \rho_0 v_0^2$,
while the energy equation reads:
\begin{eqnarray}
\frac{d}{d\xi}\left(\frac{\gamma}{\gamma -1}pv + \frac{\rho v^3}{2}\right) 
= - Q_c \nonumber
\end{eqnarray}
where $\xi$ is the distance downstream from the shock front, 
and $Q_c$ is the cooling due to the optically thin, hot plasma in CIE.

In the steady-state case when cooling is efficient, the temperature
decreases and the density increases downstream from the shock front.
The cooling layer eventually `collapses'; that is, the gas velocity
asymptotically reaches zero. It follows (see the momentum
equation) that the gas pressure does not experience large variariations in
the postshock plasma ($3/4 \leq p/\rho_0v_0^2 \leq 1$).
Thus, we adopt a constant-pressure approximation (CPA) to describe the
physical parameters of the postshock plasma. As we will see below, this
allows us to easily make use of the radiative shock model in the
analysis of the X-ray emission. Under the CPA, the energy
equation reads:
\begin{equation}
\frac{5}{2}(1+x_e)nvk\frac{dT}{d\xi} = - x_e x_H n^2 \Lambda(T)
\label{eq:energy}
\end{equation}
where the cooling term is given explicitly, $T$ is the plasma temperature, 
$k$ is the Bolzmann constant, $x_e = n_e/n$ is the relative electron 
number density, $x_H = n_H/n$ is the relative number density of
hydrogen, and $\Lambda(T)$ is the cooling function at CIE 
(e.g. \citealt{ray_76}). To facilitate the calculations, we followed a
standard approach and made use of a piece-wise power-law approximation
to the cooling curve: $\Lambda(T) = \Lambda_0 T^{\alpha}$.
In the high-temperature intervals
($10^5 \leq T \leq 4\times10^7$, and~ $T \geq 4\times10^7$)
the adopted values of the constants are 
$\Lambda_0 = 7\times10^{-19}, 3\times10^{-27} $ and 
$\alpha = -0.6, 0.5$ 
(e.g., \citealt{myas_98}).

Figure~\ref{fig:rshock} shows a comparison between the exact and the CPA
solutions for the radiative shock and illustrates that our choice
(CPA) is acceptable for the purpose of the present study. Moreover, 
as we will see below, the CPA allows an easy incorporation of the radiative 
shock model in \XSPEC which, unfortunately, 
is not the case for the exact solution of the problem.

To determine the X-ray spectrum from a steady-state, 
plane-parallel, 
radiative shock (with `surface' $A$), we must
integrate the hot plasma emissivity over a range of plasma temperatures
in the temperature-stratified postshock zone (Fig.~\ref{fig:rshock}):

\begin{eqnarray}
Sp(E) = A \int_{0}^{T_{sh}} \epsilon(E,T) 
x_e x_H n^2 d\xi 
\label{eqn:spec0}
\end{eqnarray}
where $\epsilon(E,T)$ is the specific emissivity at given photon
energy $E$ and  plasma temperature $T$ (its integral over the entire 
energy/spectral range gives the value of the cooling curve at $T$),
$\xi$ is the distance from the shock front, and $T_{sh}$ is the postshock
temperature in keV. Then, using
eq.~(\ref{eq:energy}), we can write:
\begin{equation}
Sp(E) = \Gamma_{sh} \int_{0}^{T_{sh}} \epsilon(E,T) T^{1-\alpha} d\ln T
\label{eq:spec}
\end{equation}
where for numerical convenience we integrate over  $\ln T$ instead of 
$T$, and $\Gamma_{sh}$ is the normalization parameter.
In this form,  the X-ray spectrum from a radiative shock is
well suited for modelling with \XSPEC : 
namely, we simply have a known 
power-law distribution of emission measure at various temperatures,
and we adopt an optically-thin plasma model 
(model {\it vapec} in \XSPEC)
for the plasma spectrum at temperature $T$ 
(given by the quantity $\epsilon[E,T]$).
Thus, our \XSPEC shock model have the following 
parameters: the postshock temperature $T_{sh}$, chemical abundances
and the normalization factor $\Gamma_{sh}$, which is related to the
total emission measure in the shock (some technical details are 
found in Appendix \ref{app}).

\begin{table*}
\begin{minipage}{126mm}
\caption{OB stars with {\it Chandra} Grating Spectra}
\label{tab:obs}
\begin{tabular}{rrllrccc}
\hline
\hline
\multicolumn{1}{l}{No.} &
\multicolumn{1}{c}{HD} &\multicolumn{1}{c}{Name}  &
\multicolumn{1}{c}{Type$^{a}$~~} & \multicolumn{1}{c}{Distance$^{b}$} & 
\multicolumn{1}{c}{Observation} & 
\multicolumn{1}{c}{Counts$^{c}$} & \multicolumn{1}{c}{Data$^{d}$} \\
\multicolumn{1}{c}{}  &
\multicolumn{1}{l}{} &\multicolumn{1}{c}{}  &
\multicolumn{1}{c}{} & \multicolumn{1}{c}{(pc)} &
\multicolumn{1}{c}{IDs} & 
\multicolumn{1}{c}{(MEG)} & \multicolumn{1}{c}{} 
 \\
\hline
\hline
1 &150136~~ &                 &  ~~O3     & 1300~~~ &  
2569  &   8569 & 8 \\
2 &66811~~  & $\zeta$ Pup     &  ~~O4 I   &  450~~~ & 
~640   &   17435~~ & 1, 11 \\
3 &37022~~  & $\theta^1$ Ori C&  ~~O4-6p  &  450~~~ &
3, 4   &   28719; 13706 & 3, 6, 7, 11 \\
4 &149757~~ & $\zeta$ Oph     &  ~~O9V    &  140~~~ & 
2571, 4367   & 5915 &   \\
5 &36486~~  & $\delta$ Ori A  &  ~~O9.5 II&  501~~~ & 
~639   &   6116 & 4, 11 \\
6 &37742~~  & $\zeta$ Ori A   &  ~~O9.7 Ib&  501~~~ &
~610   &   9220 & 10, 11 \\
7 &37128~~  & $\epsilon$ Ori  &  ~~B0Iab: &  412~~~ &
3753   &   6811 &   \\
8 &149438~~ & $\tau$ Sco      &  ~~B0.2 V &  132~~~ &
~638   &   16838~~ & 2, 11 \\
9 &206267A&                 &  ~~O6.5 V & 800~~~ & 
1888, 1889  & 1597 & 5, 11 \\
10 &47839~~  & 15 Mon          &  ~~O7 V   & 767~~~ & 
5401, 6247  & 1631 &   \\
11 &57061~~  & $\tau$ CMa      &  ~~O9 II  & 1480~~~ & 
2525, 2526   & 1470 & 5, 11 \\
12 &37043~~  & $\iota$ Ori     &  ~~O9 III &  440~~~ & 
599, 2420   & 4934 & 11 \\
13 &37468~~  & $\sigma$ Ori AB &  ~~O9.5 V & 353~~~ & 
3738  &   1893 & 9 \\
14 &111123~~ & $\beta$ Cru     &  ~~B0.5 III&  110~~~ &
2575   &   1830 & 11 \\
15 &116658~~ & Spica           &  ~~B1 III-IV&  80~~~ & 
4509   &   3041 &   \\
\hline
\hline
\end{tabular}

\medskip
$^{a}$ Spectral type is from Wojdowski \& Schulz (2005)
except for HD~150136, $\zeta$ Oph, $\epsilon$ Ori, 15 Mon,
$\sigma$ Ori AB, and Spica whose spectral type is from SIMBAD.

$^{b}$ Values of the distance are taken
from Wojdowski \& Schulz (2005) except for 
HD~150136 (Herbst \& Havlen 1977), and
$\zeta$ Oph, $\epsilon$ Ori, 15 Mon, $\sigma$ Ori, 
and Spica (from the Jim Kaler's `STARS' web site: 
http://www.astro.uiuc.edu/$\sim$kaler/sow/sowlist.html).

$^{c}$ The total number of counts in the {\it Chandra} MEG 
background-subtracted spectra. 
Thanks to the X-ray brightness of $\theta^1$ Ori C, the two
observations were analyzed separately. The data for Spica were
obtained with the {\it Chandra} LETG. 

$^{d}$ References to the articles where the X-ray spectra of the stars 
are discussed in detail: 
(1) \citet{cass_01};
(2) \citet{co_03};
(3) \citet{ga_05};
(4) \citet{mi_02};
(5) \citet{schulz_03};
(6) \citet{schu_00};
(7) \citet{schu_03};
(8) \citet{sk_05};
(9) Skinner et al. (2007, in preparation);
(10) \citet{wa_00};
(11) \citet{woj_05}.

\end{minipage}
\end{table*}

\subsection{Global Model}
\label{subsec:global}
Our global model assumes that X-rays are produced in radiative shocks
which are randomly (and uniformly) distributed 
in the stellar wind, most likely in the acceleration zone. Therefore,
the spectral lines in the total (integrated) X-ray spectrum from shocks 
are broadened by the bulk (wind) gas velocity, but are not shifted without
including absorption. 
Qualitative physical considerations, also demonstrated by numerical 
modelling (e.g., \citealt{ow_01}), show that the X-ray absorption  
effects in the stellar wind result in blueshifted spectral lines with 
asymmetric line profiles.

To describe this physical situation we have
constructed a new XSPEC model, whose technical aspects are given
in Appendix \ref{app}. The main features of the model are:
(1) shocks are characterized by their
postshock temperature and the distribution of the total emission measures 
in the shocks is determined from the Chebyshev polynomial algorithm, as
is used in the standard \XSPEC model $c6pvmkl$ \citep{le_89}.
We note that a similar approach was successfully adopted in fitting
the X-ray emission from an ensemble of adiabatic shocks \citep{zh_06}.
(2) The basis vectors for X-ray emission from the shock ensemble
are those from the radiative shock model incorporated in \XSPEC 
(see \S~\ref{subsec:shock}). (3) All shocks have the same elemental 
abundances and share the same spectral line shifts. 
Thus, the basic model parameters are: the distribution of emission
measure of radiative shocks; chemical abundances, and the global line 
shift for the spectrum.

The profile asymmetry is handled through two (half)Gaussians each
approximating the blue- and red-halves of the line profile.
The Gaussian widths can vary with wavelength (in the samer manner as in 
the \XSPEC model {\it gsmooth}: $\Delta E \propto E^{\beta}$). 
Although the goal of our approach
is just to obtain best fits to the line profiles over the
entire spectrum, it is worth noting that the wavelength dependence of 
the line width represents the physical notion that faster shocks produce 
higher-temperature plasma whose emission is dominated by lines at 
shorter wavelengths.

\section{X-ray Data} 
\label{sec:obs}
For our study, we selected a sample of OB stars with available
gratings observations in the  {\it Chandra} data archive. 
We extracted first-order spectra for each object according to the
procedure described in the Science Threads for Grating Spectroscopy in
the {\small CIAO} 3.3 \,\footnote{Chandra Interactive Analysis of 
Observations ({\small CIAO}), 
http://cxc.harvard.edu/ciao/} data analysis software.
The ancillary response
functions for all spectra were generated using the Chandra
calibration database {\small CALDB} v3.2. Table~\ref{tab:obs} gives some 
basic information about the stars of the sample and the X-ray observations.
Our analysis is based on the MEG spectra to take advantage of their
higher sensitivity compared to the HEG spectra. One object (Spica) was 
observed with the {\it Chandra} LETG. 

Depending on the quality of the data (the
total number of photons), the MEG spectra were rebinned to have
between 15 and 30 counts per bin. The spectra near the intercombination 
and the forbidden lines in the helium-like triplets (Si XIII, Mg XI, 
Ne IX and O VII) were rebinned so that these two lines fall
into one large bin. This technical approach is aimed at improving
the quality of the fit when optically thin plasma models (as those
available in \XSPEC) are used that
do not take into account the effects of a strong external UV field.
In the case of hot massive stars, such a UV field is present and it 
may alter the ratio of the two lines in the
helium-like triplets, although it does not change their sum (the total
emissivity/intensity). Thus, this technical approach allows to
preserve the piece of information available in the helium-like
triplets that can contribute in constraining some plasma
characteristics (e.g., plasma temperature, abundances) which come out
from the spectral fits.

\section{Results}
\label{sec:res}
The X-ray spectra (MEG) of the stars from our sample (see
Table~\ref{tab:obs}) were fitted with the global model that assumes a
distribution of radiative shocks in the stellar wind
(\S~\ref{subsec:global}). To determine the elemental abundances, we
varied only those having strong emission lines in the observed spectum
of a given object. These usually include N, O, Ne, Mg, Si and Fe, while
for some objects C and S were also varied. For the abundance of all the other 
elements, the solar value was adopted \citep{an_89}.
Finally, to derive the X-ray absorption column density toward each object,
we made use of \citet{mo_83} cross sections 
(\XSPEC model {\it wabs}).

\begin{figure*}
\includegraphics[width=10in, height=180mm, angle=-90]{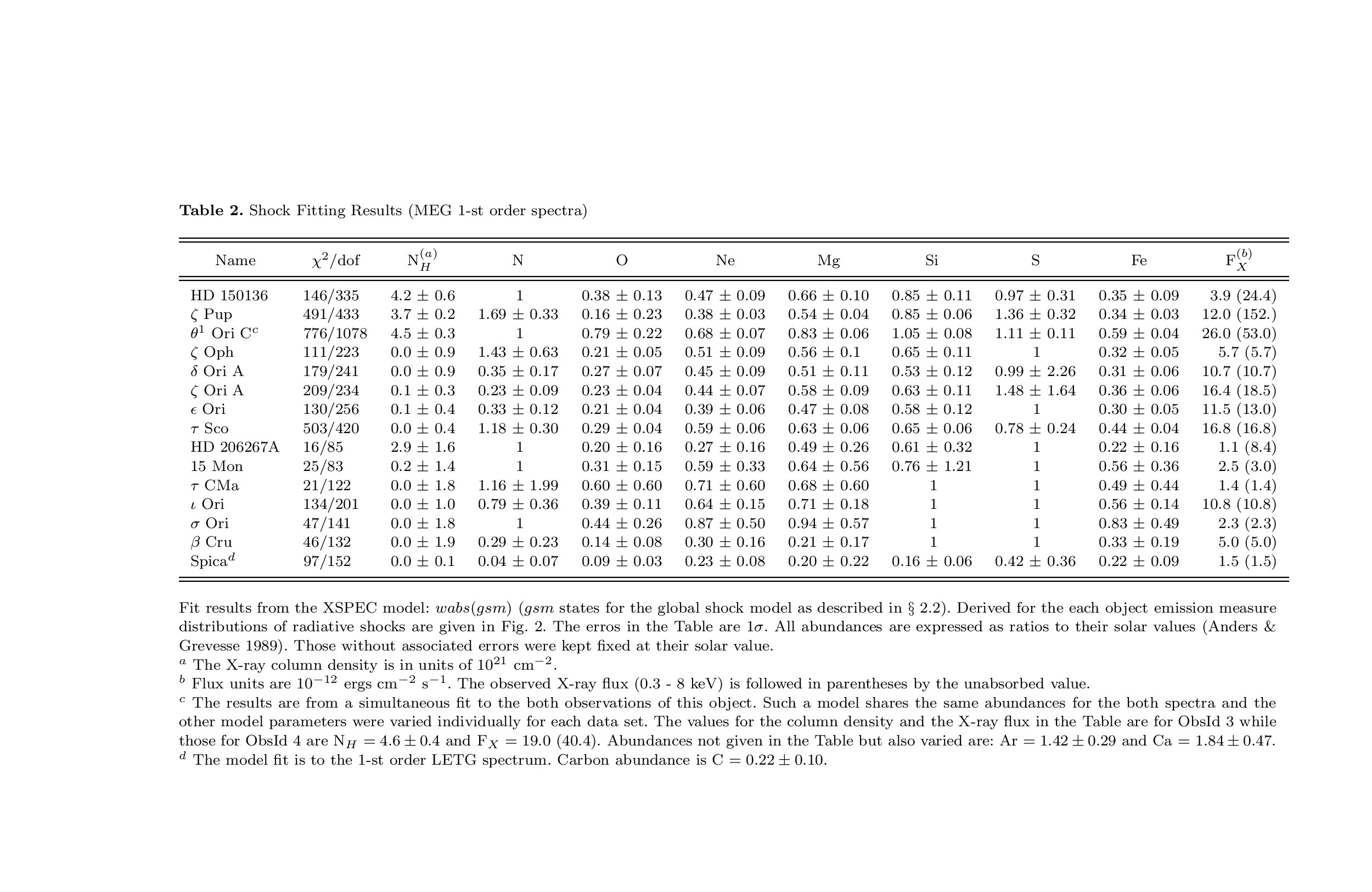}
\end{figure*}

\begin{figure*}
\includegraphics[width=2in, height=1.5in]{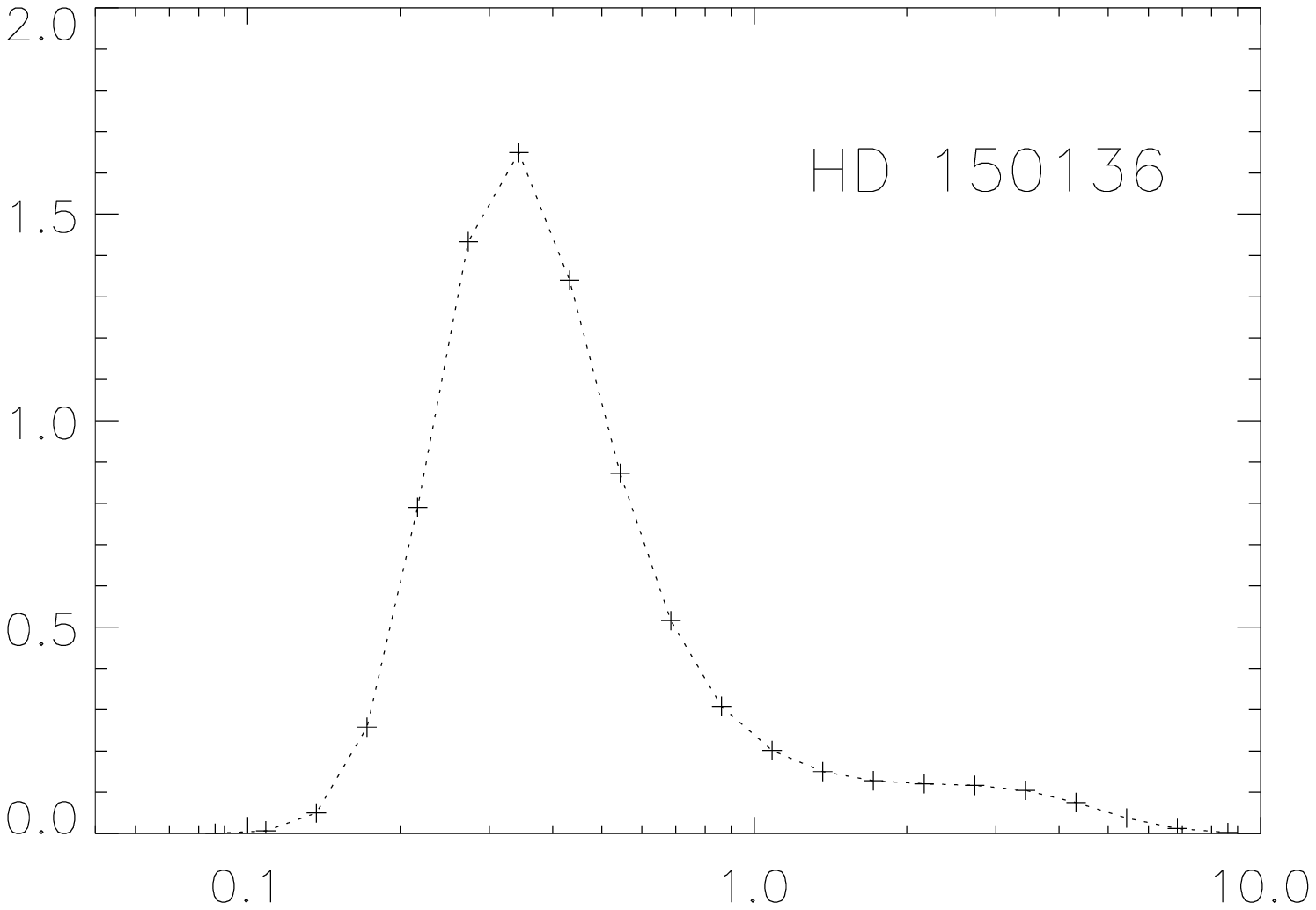}
\includegraphics[width=2in, height=1.5in]{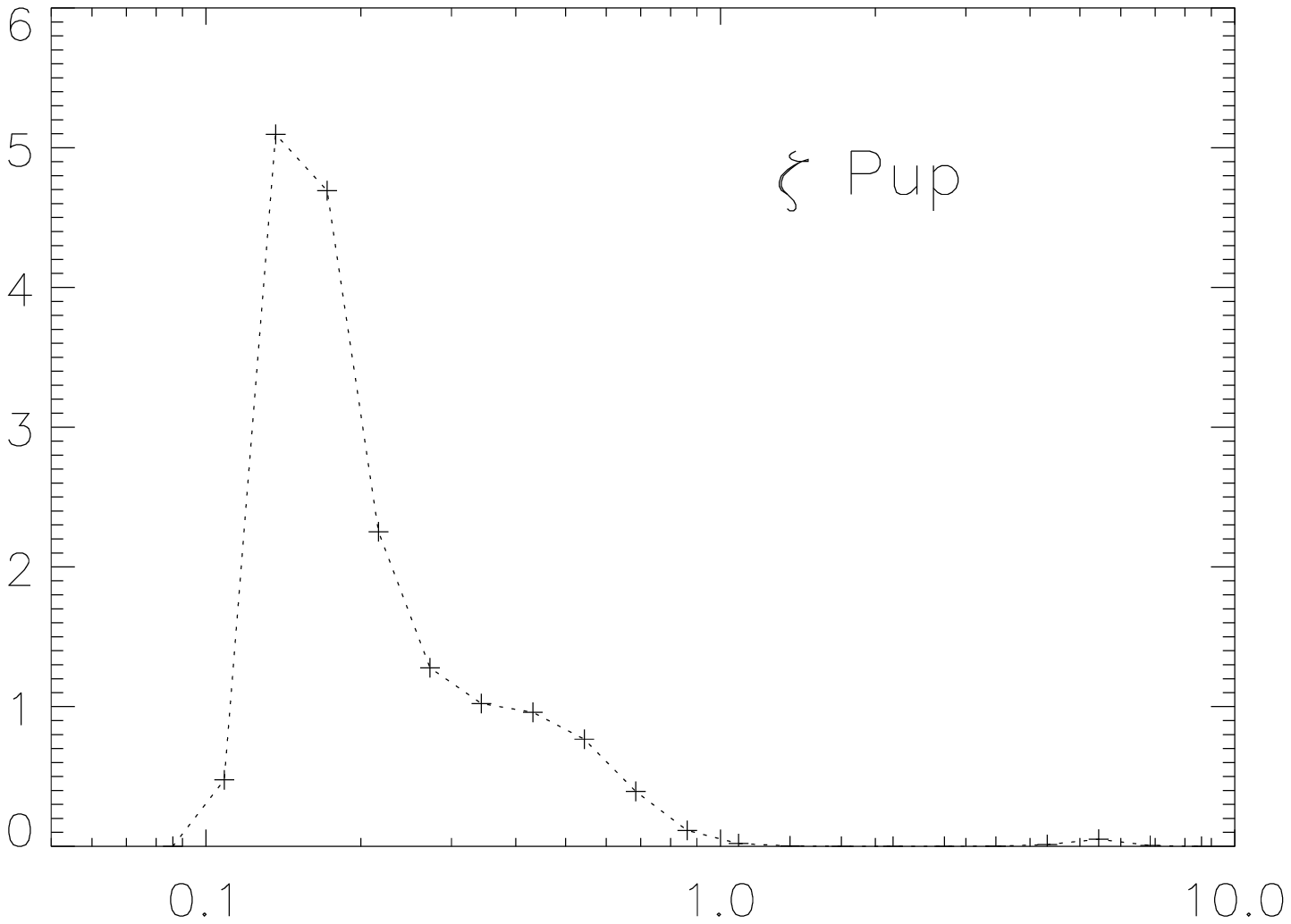}
\includegraphics[width=2in, height=1.5in]{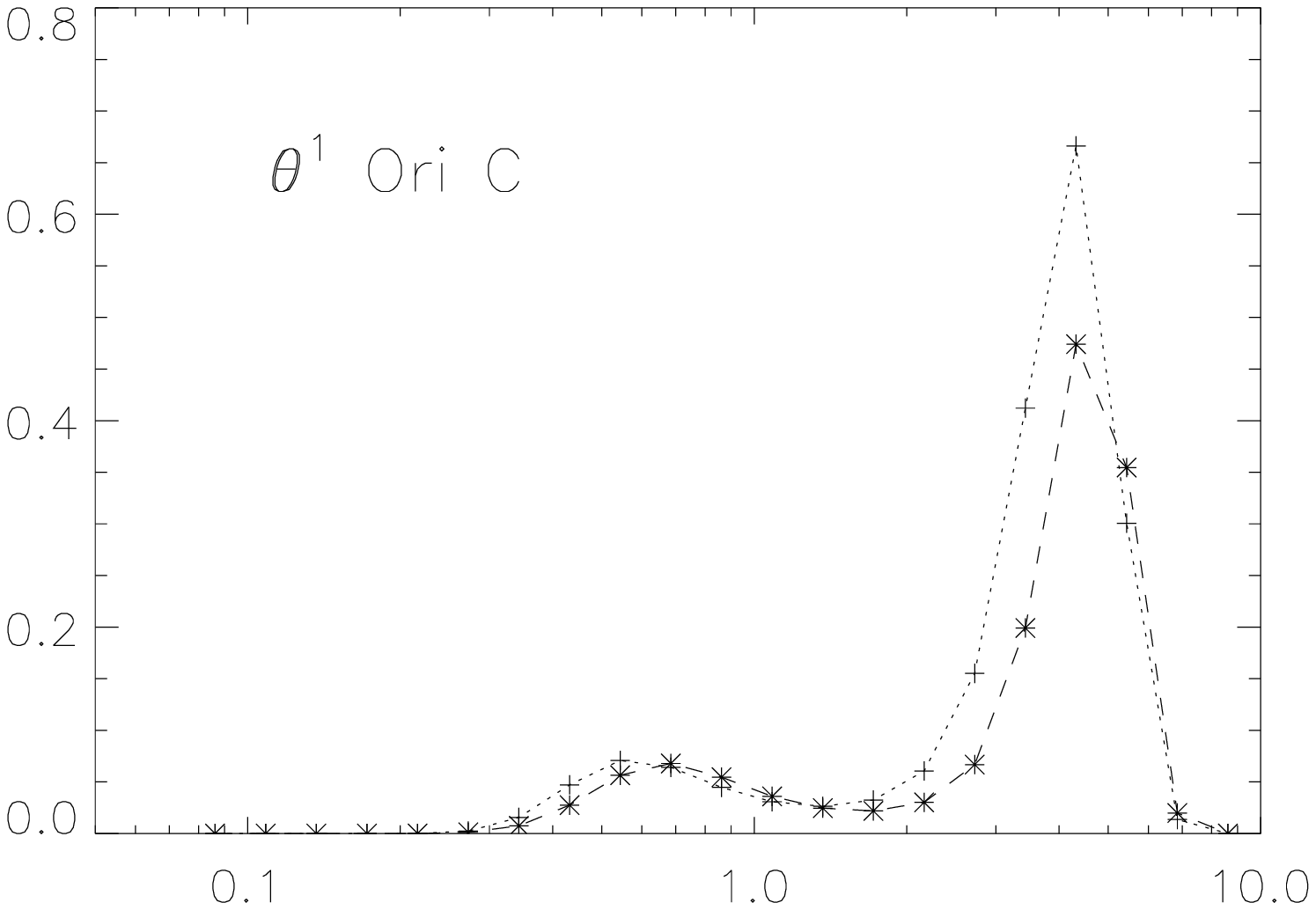}
\includegraphics[width=2in, height=1.5in]{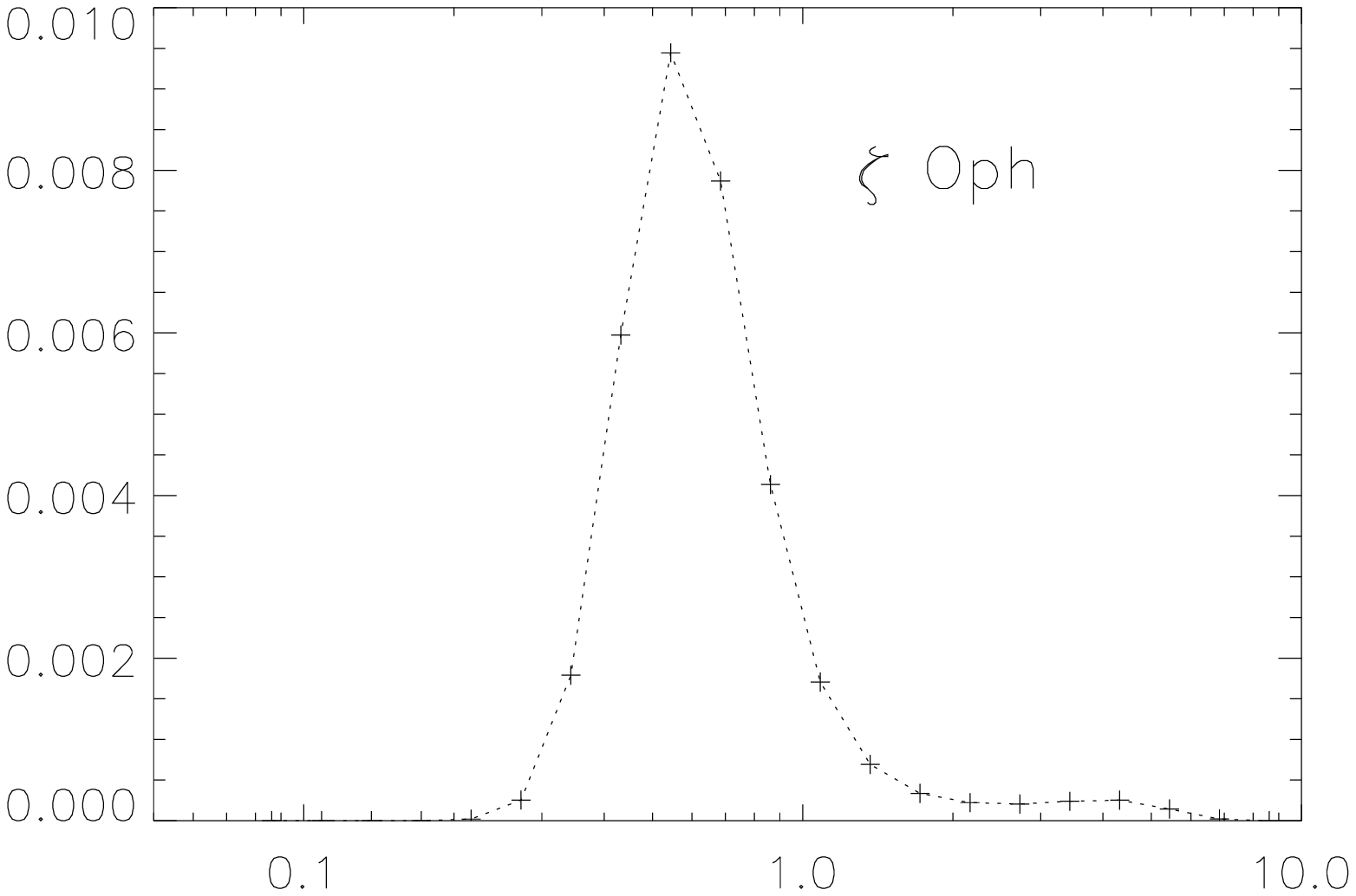}
\includegraphics[width=2in, height=1.5in]{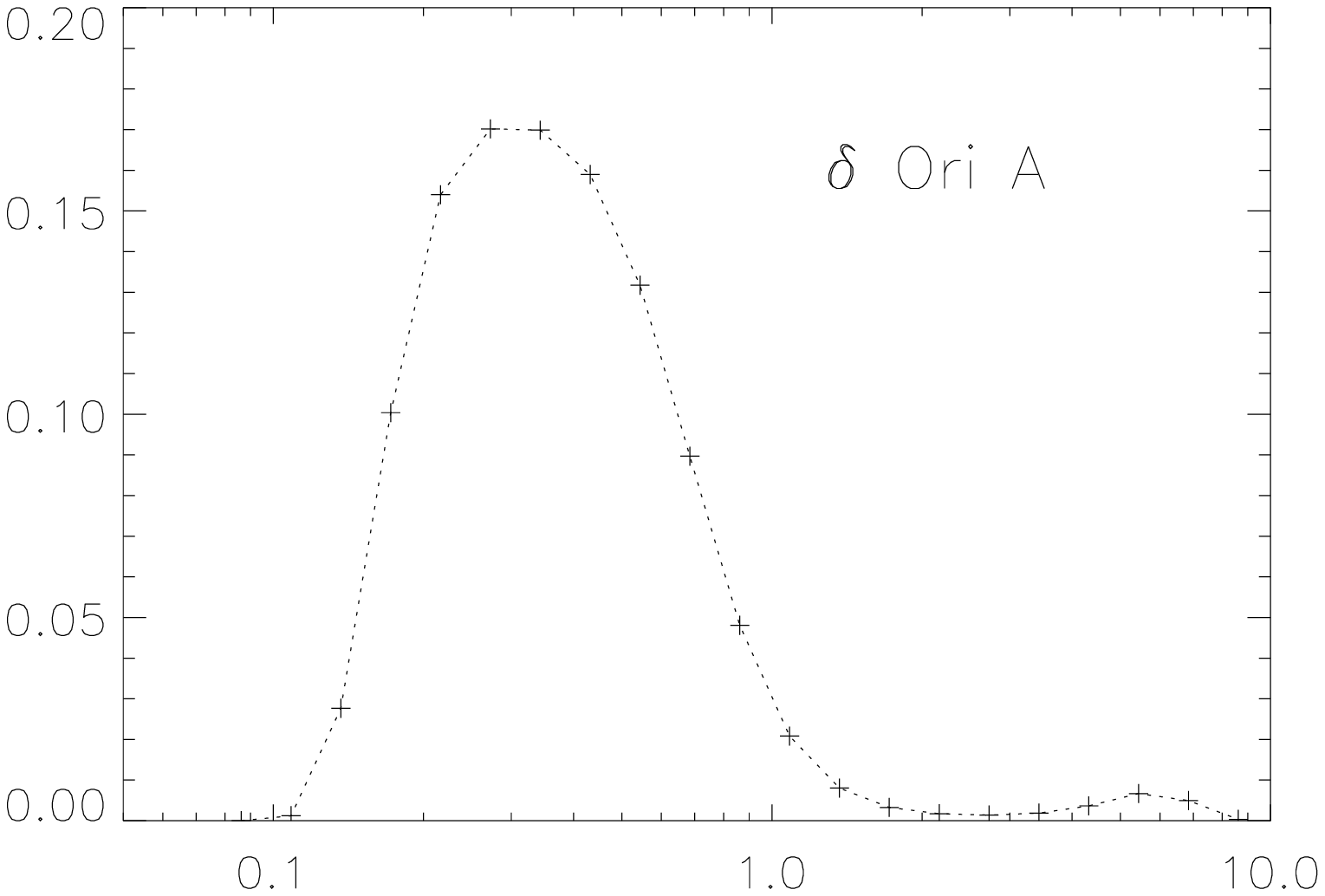}
\includegraphics[width=2in, height=1.5in]{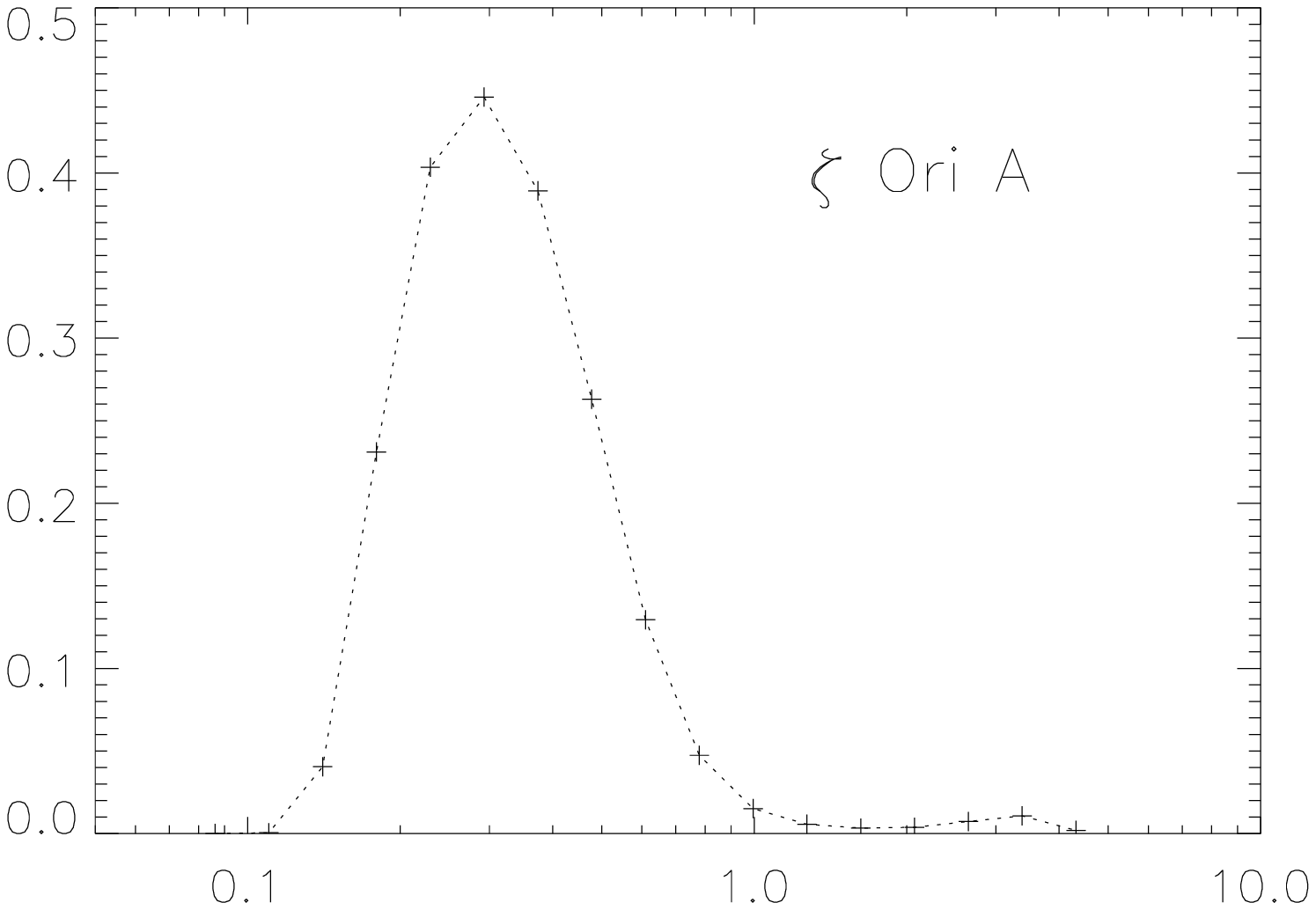}
\includegraphics[width=2in, height=1.5in]{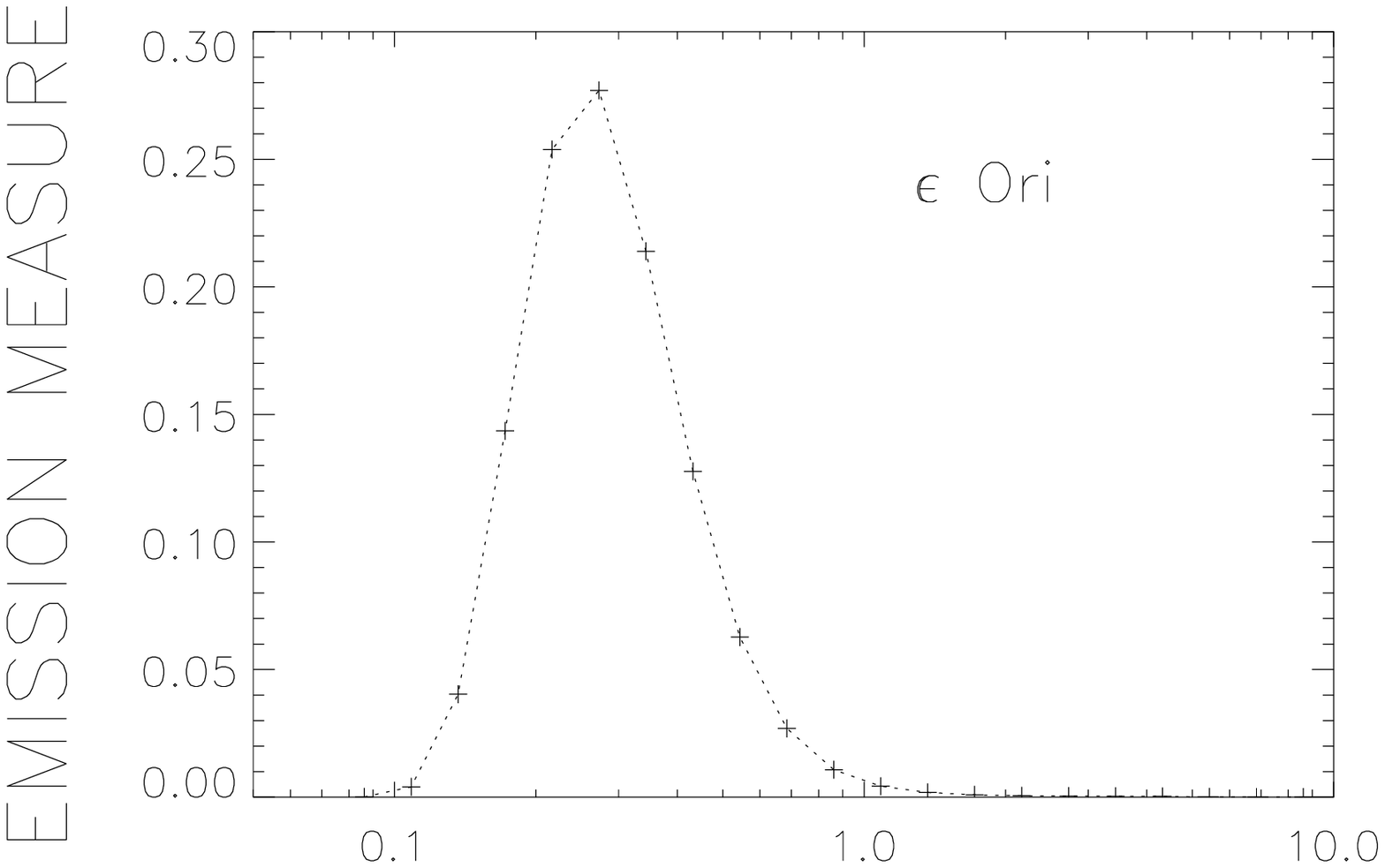}
\includegraphics[width=2in, height=1.5in]{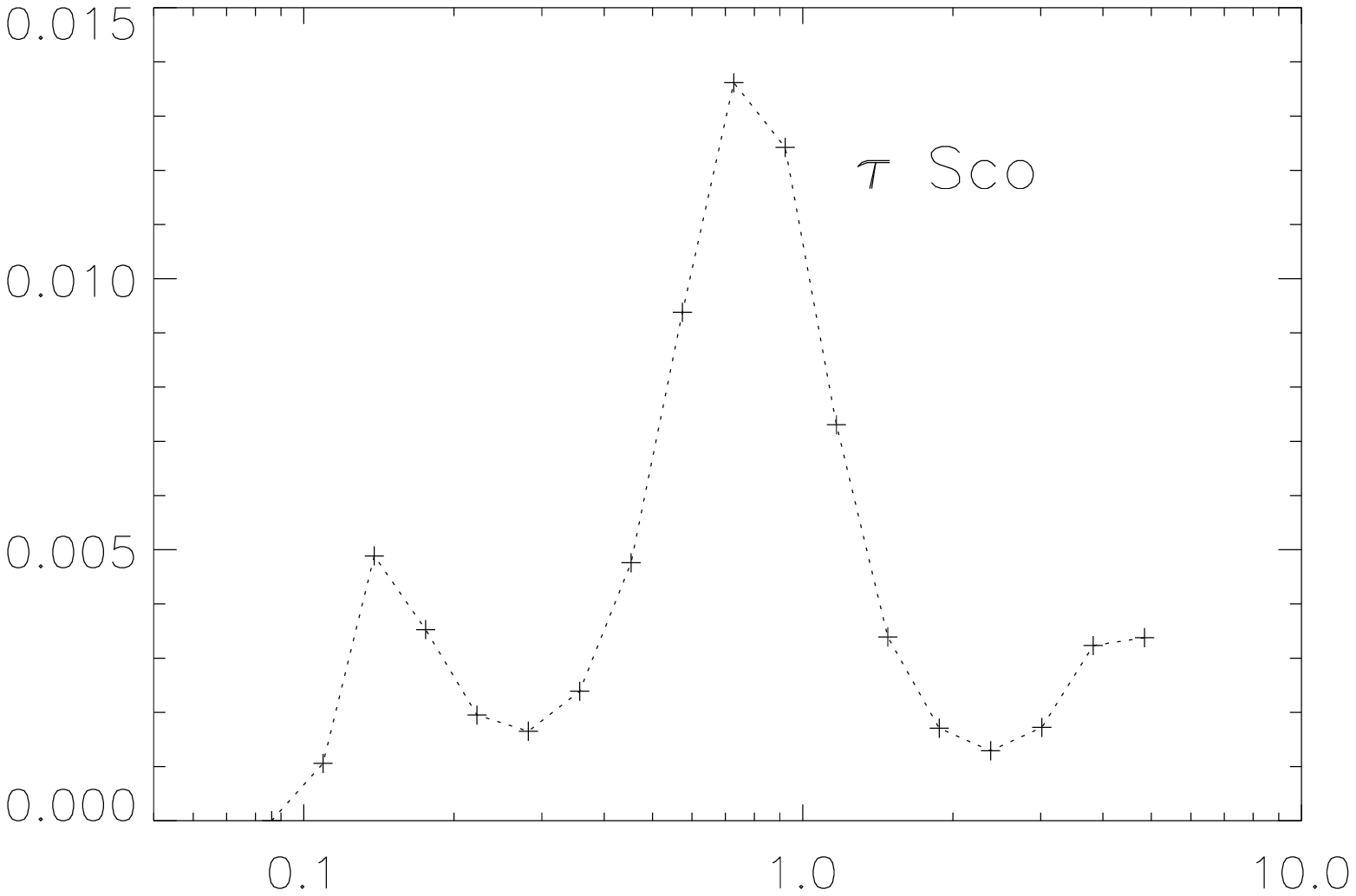}
\includegraphics[width=2in, height=1.5in]{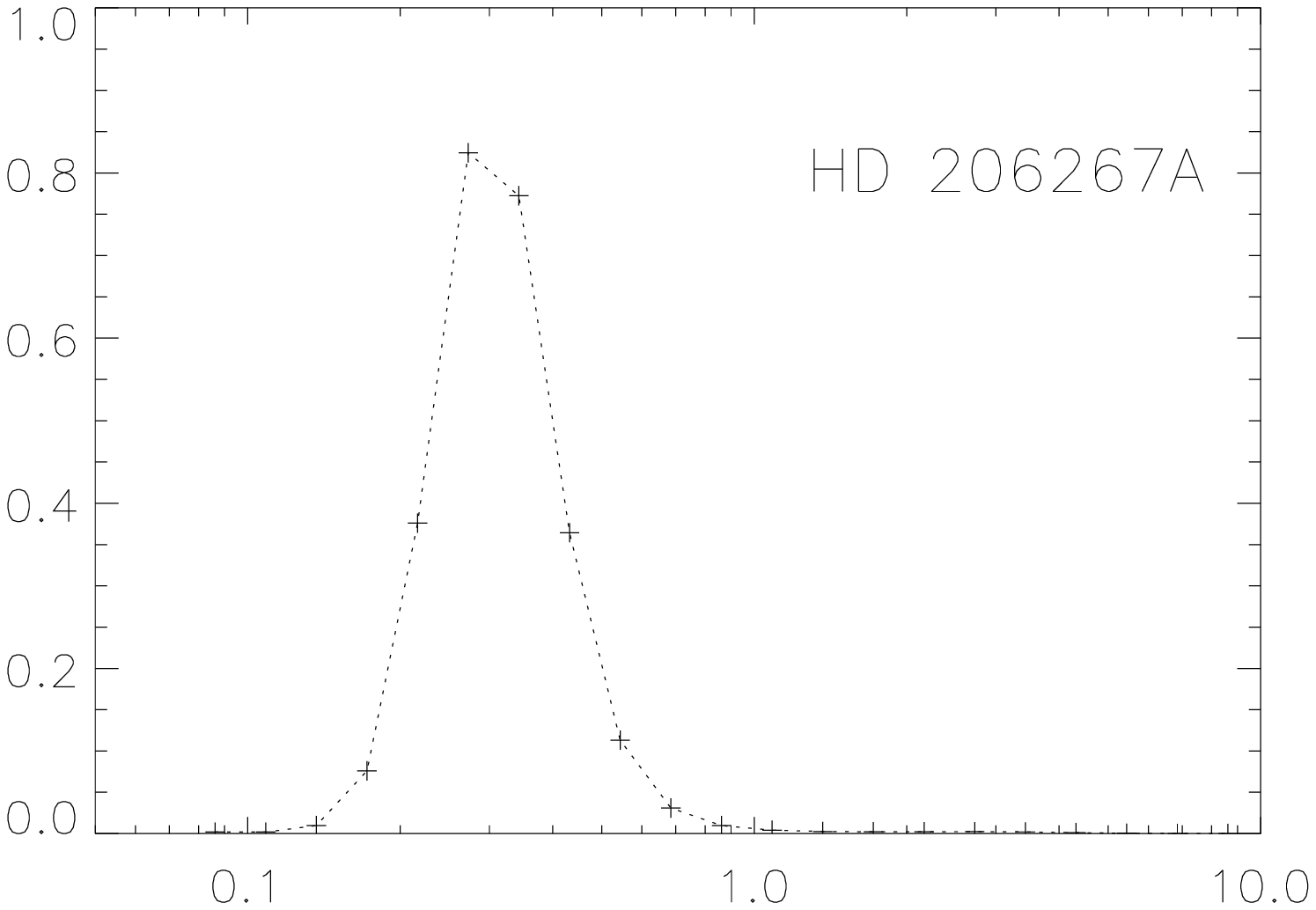}
\includegraphics[width=2in, height=1.5in]{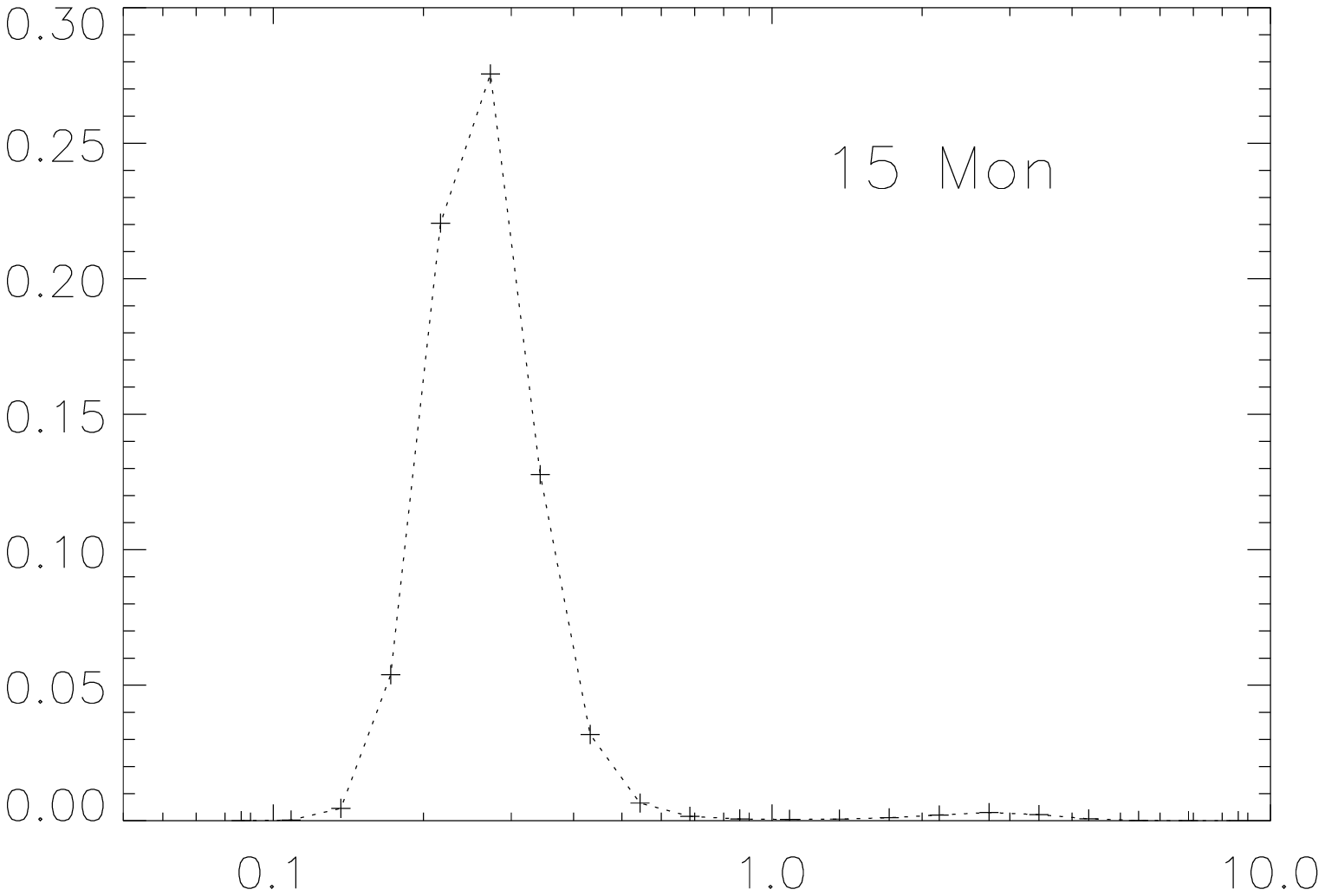}
\includegraphics[width=2in, height=1.5in]{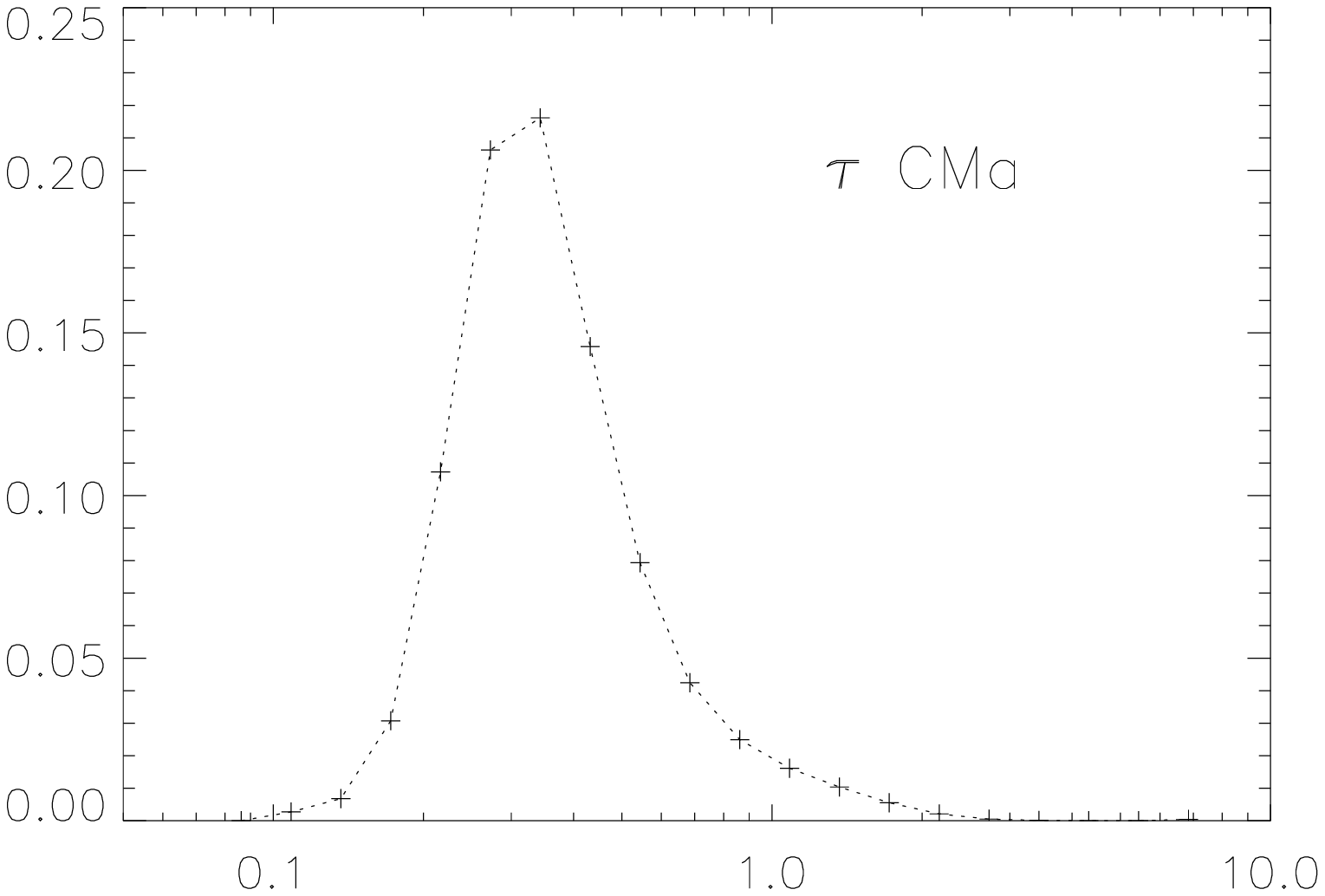}
\includegraphics[width=2in, height=1.5in]{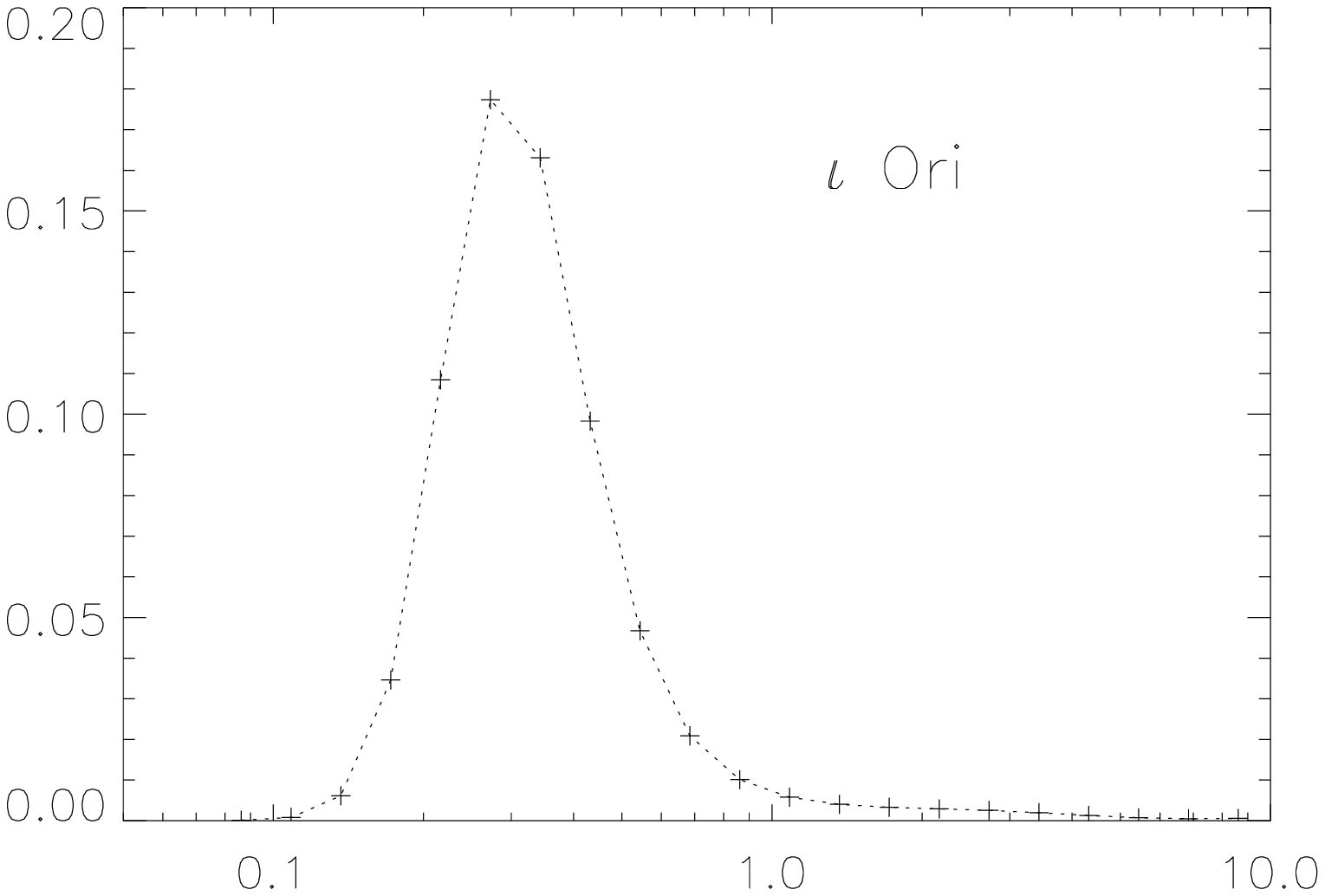}
\includegraphics[width=2in, height=1.5in]{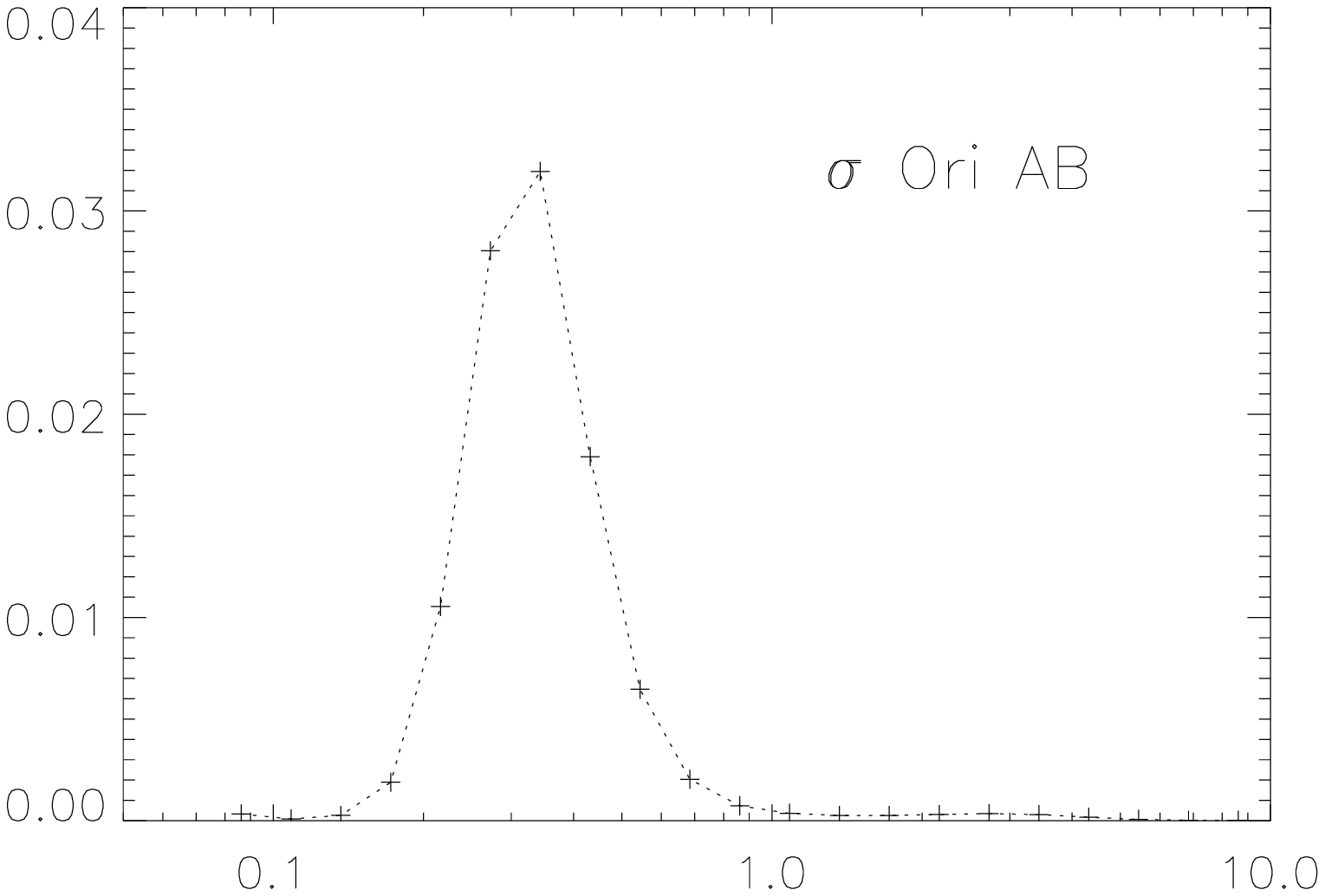}
\includegraphics[width=2in, height=1.5in]{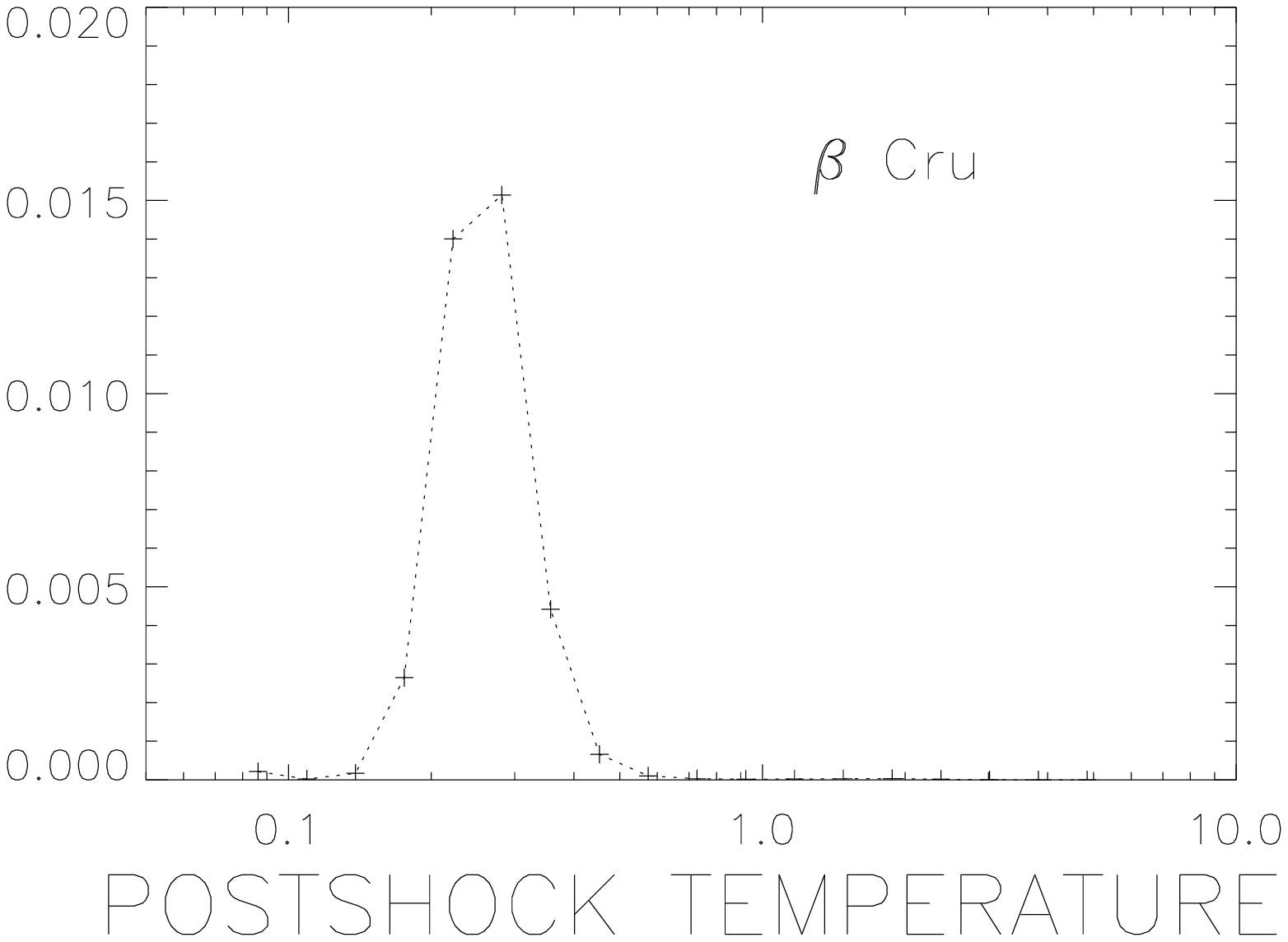}
\includegraphics[width=2in, height=1.5in]{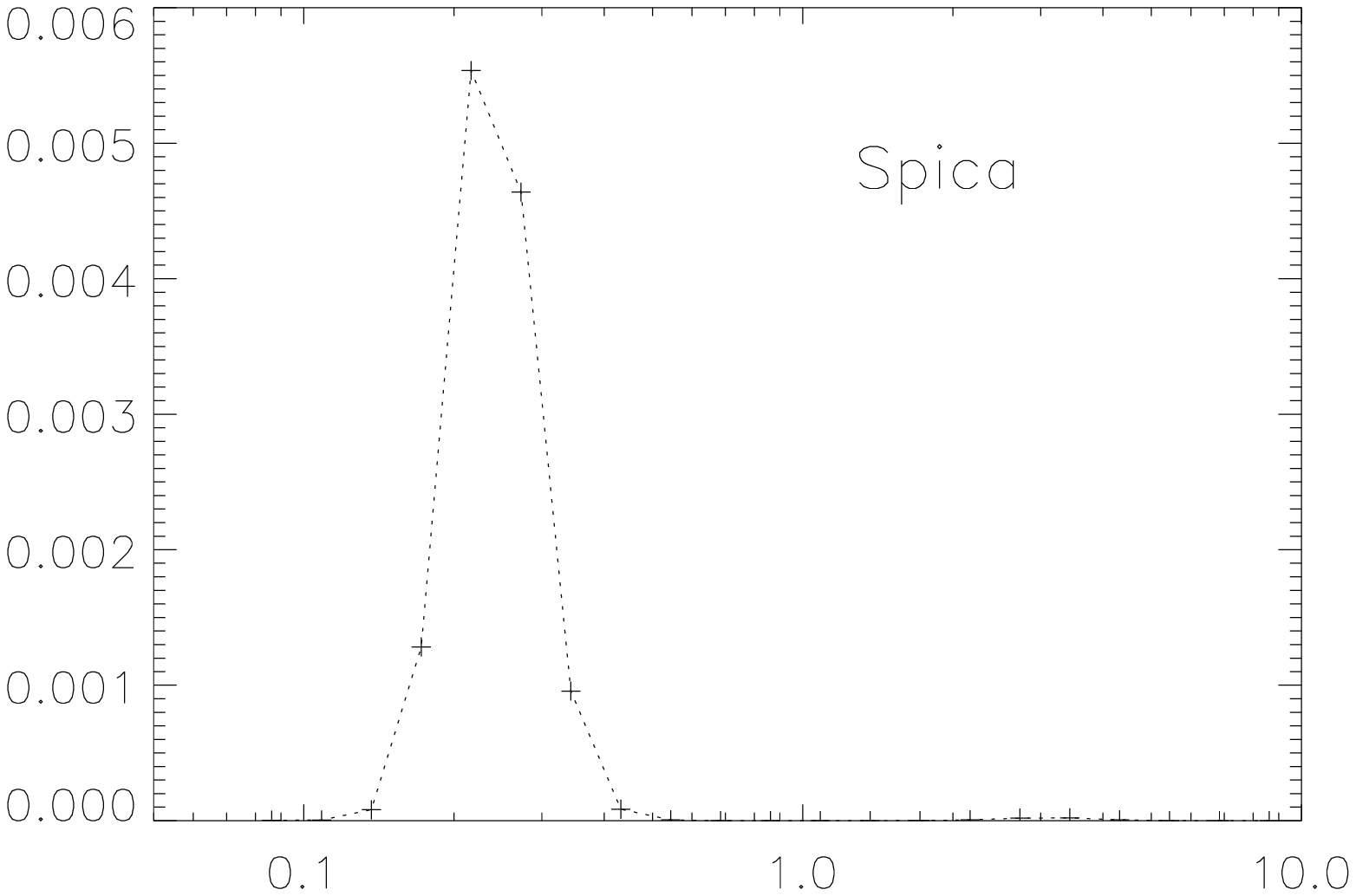}
\caption{Emission measure distribution of radiative shocks in the OB
stars of our sample. 
{\bf Horizontal axes} are the immediate postshock temperature in keV;
{\bf vertical axes} are the emission measure in units of 
$10^{56}$~cm$^{-3}$.
In the case of $\theta^1$ Ori C, the results for ObsId 3 and ObsId 4
are shown by dotted and dashed lines, respectively.
}
\label{fig:dem}
\end{figure*}

\begin{figure*}
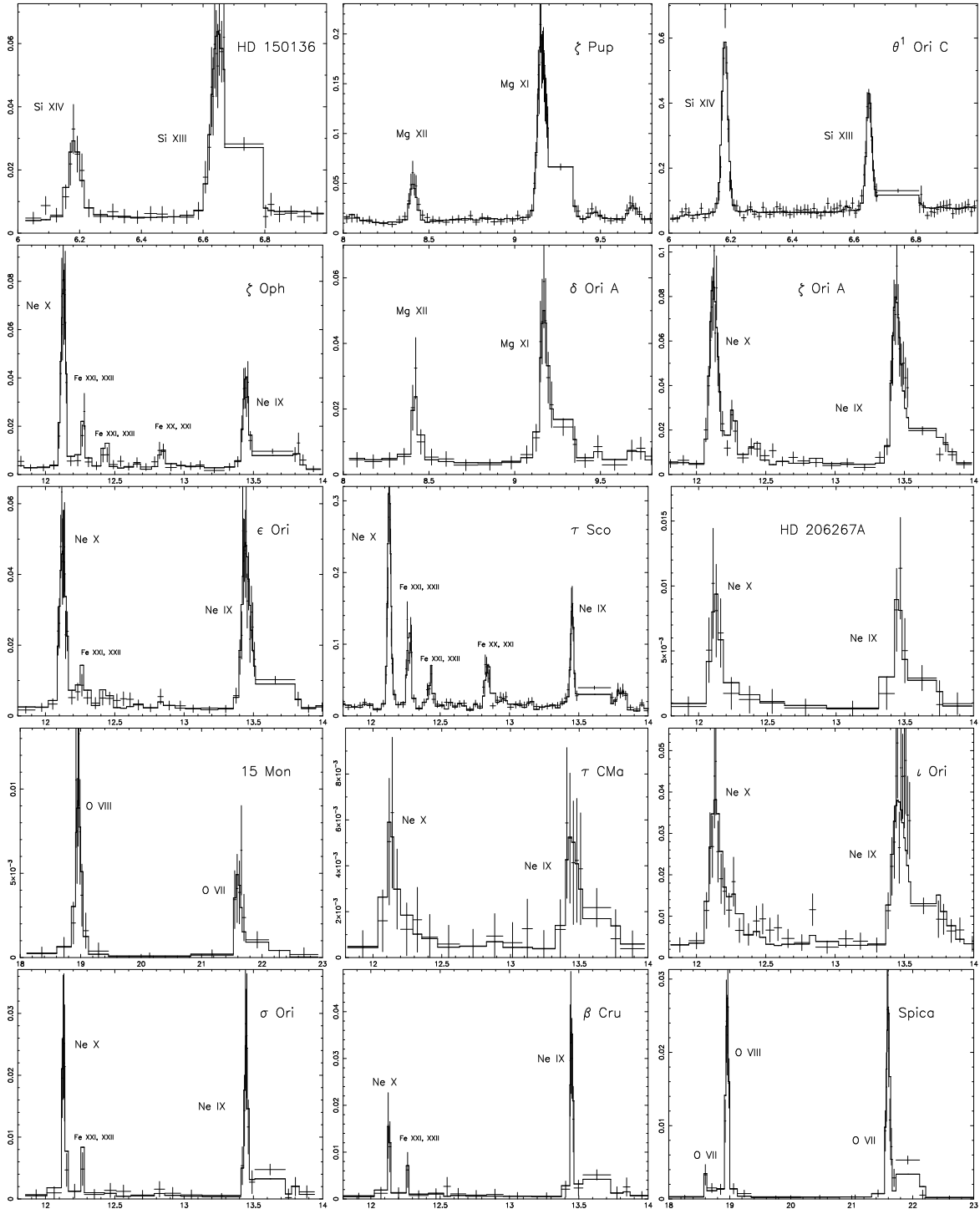

\includegraphics[width=1.5in, height=2in,angle=-90]{fig3a.eps}
\includegraphics[width=1.5in, height=2in,angle=-90]{fig3b.eps}
\includegraphics[width=1.5in, height=2in,angle=-90]{fig3c.eps}
\includegraphics[width=1.5in, height=2in,angle=-90]{fig3d.eps}
\includegraphics[width=1.5in, height=2in,angle=-90]{fig3e.eps}
\includegraphics[width=1.5in, height=2in,angle=-90]{fig3f.eps}
\includegraphics[width=1.5in, height=2in,angle=-90]{fig3g.eps}
\includegraphics[width=1.5in, height=2in,angle=-90]{fig3h.eps}
\includegraphics[width=1.5in, height=2in,angle=-90]{fig3i.eps}
\includegraphics[width=1.5in, height=2in,angle=-90]{fig3j.eps}
\includegraphics[width=1.5in, height=2in,angle=-90]{fig3k.eps}
\includegraphics[width=1.5in, height=2in,angle=-90]{fig3l.eps}
\includegraphics[width=1.5in, height=2in,angle=-90]{fig3m.eps}
\includegraphics[width=1.5in, height=2in,angle=-90]{fig3n.eps}
\includegraphics[width=1.5in, height=2in,angle=-90]{fig3o.eps}
\caption{Portion of the background-subtracted spectra near some strong
emission lines. The overlaid best-fit model is also shown (solid line). 
{\bf Horizontal axes} are the observed wavelength (\AA); 
{\bf vertical axes} are the flux density (photons s$^{-1}$ \AA$^{-1}$).
}
\label{fig:fit}
\end{figure*}

Table~2 presents some basic results from the fits to the spectra 
of the studied stars: the total $\chi^2$~and the degrees of freedom
for the fit (2nd column); the 
X-ray absorption column density (3rd column); 
the individual elemental abundance relative to the solar value (columns 4-10); 
the observed and unabsorbed X-ray flux (last columns). 
And the corresponding distributions of emission measure of the radiative 
shocks in the stellar winds are shown in Figure~\ref{fig:dem}. 
The quality of the global  fits in the vicinity of some strong emission 
lines is illustrated in Figure~\ref{fig:fit}.
A few properties are worth noting. 

There is an indication 
that the metal abundances are subsolar in the sample of OB stars with
gratings spectra and this trend is well demonstrated by the iron abundance
being a factor of 2-3 below the solar value. Nitrogen might be considered
an exception but we note 
that the poor quality of the data in the soft part of the X-ray spectra, where
the nitrogen lines are found, prevents a firm conclusion.
The values of the column density of the X-ray absorbrption material are
consistent with those derived from the visual extinction to each object,
using data from \citet{berg_96} and the \citet{go_75} conversion formula.
The only object showing a clear sign of excess absorption with respect to
its interstellar value is $\zeta$~Pup which may suggest that wind
absorption effects play an important role in this object.


We see that most stars display the same
distribution of emission measure of the radiative shocks,
peaked at $\sim$0.1-0.4 keV and with a very low level of high temperature 
tail extending above 1.0 keV. Thus, in most cases X-rays are 
produced in low velocity shocks.  This finding is consistent with the 
radiation-driven instabilities scenario.
The two most notable exceptions are $\theta^1$ Ori C and $\tau$ Sco.
In the former case, the main peak is at energy $\sim$4-5 keV with a
weaker component at $\sim$0.8 keV. 
The case of $\tau$ Sco is more peculiar since there appear to be three
peaks: one at very low energies, the main one at $\sim$0.7 keV, and some
emission above $\sim$3 keV. 
One could argue that the hot tail in the distribution of emission
measure in these objects is due to the MCWS while the cool component comes
from the RDI shocks. 
For example, the shape of the emission measure distribution in 
$\theta^1$ Ori C is stable between the two data sets (taken at two
different times) and only the 
total amount of the hot gas changed mostly in the high-temperature
component (Fig.~\ref{fig:dem}). We note that this finding is in accord 
with the previous analysis of {\it Chandra} data 
(\citealt{schu_03}; \citealt{ga_05}).
Such a different behaviour of the hot and cool components might be
an indication of their different origin, as mentioned above. 
However, it is necesary to obtain more data also on $\tau$ Sco, 
in order to establish whether such a behaviour is typical only for
one object or has much more general validity.
It should be noted that there is a third object in our sample of
OB stars, $\zeta$ Oph, whose distribution of radiative shocks 
is relatively `hot' and it peaks at $\sim 0.5-0.6$ keV.

The presence of considerably hotter plasma (therefore,
fast velocity shocks), as definitely seen in $\theta^1$ Ori C, $\tau$ Sco
and $\zeta$ Oph, in combination with the relatively narrow lines
observed in their spectra could be assumed as a clear manifestation of the
magnetically-confined wind effects. It was recently proposed by 
\citet{schu_03} that the efficiency of the MCWS mechanism correlates with the
age of the massive star. The relative youth of the first two objects
(0.3 and $\sim1$~Myr, see their Table~4) and the small kinematic 
age of $\zeta$ Oph ($\sim1$~Myr, \citealt{hoo_01}) indeed support this 
suggestion.

\section{Discussion}
\label{sec:dis}
The radiative shock model introduced in this study
allows us to deduce some important information on the global
physical state of the X-ray emitting region of massive stars. As
one can see from eq. (A1), the normalization factor of the flux,
$\Gamma_{sh}$,
depends on the shock temperature ($T_{sh}$), 
stellar wind parameters ($\dot{M}_6$, $v_{1000}$), 
and distance to the object in kpc ($d_{kpc}$):
$
\Gamma_{sh} = 2.444~\delta \left( \frac{\dot{M}_6}{v_{1000}} \right)
       \frac{\sqrt{T_{sh}}}{d^{~2}_{kpc}}
$, as $\delta$ gives the 
effective surface of the shock.
Thus, from the results of the fits to the X-ray spectra, namely,
using the $\Gamma_{sh}$ value
and knowing the distance to each star,
we can deduce some useful information on the structure of the stellar
winds. To do so, 
we introduce a new
quantity, $\delta \left( \frac{\dot{M}_6}{v_{1000}} \right)$,
that is related to a specific shock, or to a group of shocks with 
the same velocity. 
We will refer to this quantity as the specific stellar `cloudiness'. 

\begin{figure*}
\includegraphics[width=5in, height=3.9in]{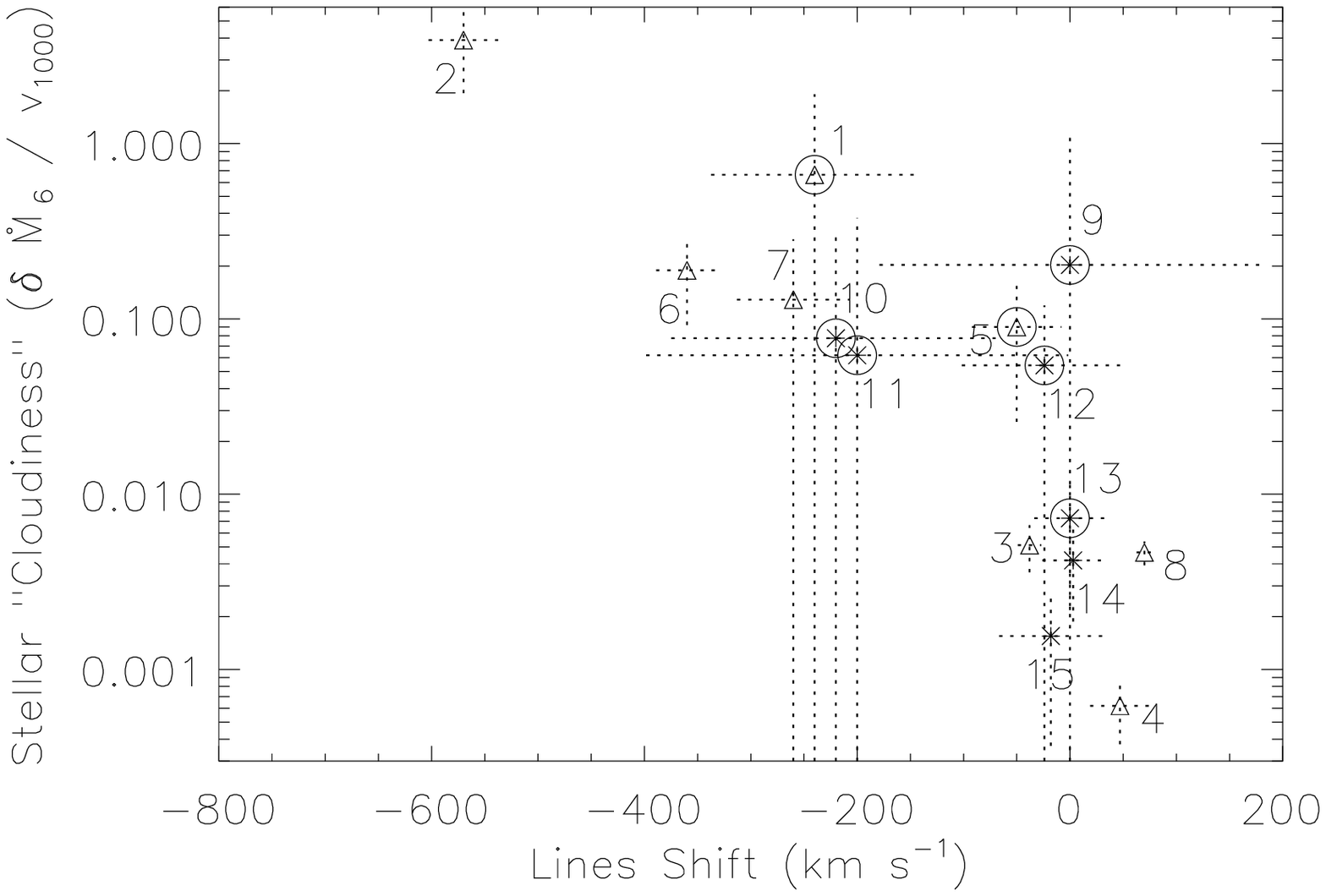}
\caption{
Total stellar `cloudiness' 
vs. global line shift for the OB stars of the sample.
The error bars (dotted lines) are $1\sigma$ errors from the fit
(the errors of the total stellar `cloudiness' follow from the
$1\sigma$ errors
for the total emission measure as derived in the fit, and the
uncertainties of the distance to the studied object are not taken into
account in the error propagation).
The objects with good photon statistics (total number of counts in
the spectrum higher than 5000) are marked by triangles.
Data points within circles are binaries or multiple systems
(as identified in SIMBAD). Stars are identified by numbers as in
Table~\ref{tab:obs}.
The data point for $\theta^1$ Ori C (number 3)
is the average of the two observations.
}
\label{fig:cloudy}
\end{figure*}

In our picture, the winds of hot massive stars are
clumpy and the clumps are likely generated by radiation-driven 
instabilties, particularly efficient in the inner acceleration zone.
The instabilities develop into strong shocks which are the source
of X-rays. Since the shocks are radiative,
the gas density in the cooling zone is much larger (probably
by a factor of a few tens and more) than the local density of the
smooth wind. Away from the acceleration zone, these shock
structures (clumps, complete or fractional shells) expand and 
cool down even further, but the density contrast in their `tail' zone 
will remain appreciably high. 
Eventually,  the clumps become `cold' 
clouds in the X-ray sky of the hot massive star, and in a steady-state
picture there will be a correlation between their characteristics:
e.g., the higher the number 
of the hot clumps, the more numerous the `cold' clouds; 
the higher the clumps' covering factor, the higher the chance for X-ray
absorption by the `cold' clouds.
In such a picture, the X-rays emitted from a particular shock will
likely experience some absorption by the cold gas in its cooling zone, 
and there is a chance for more X-ray absorption to take place in the
`cold' tails of the other shocks and the 'cold' clouds as well.
Although the exact situation depends on many specific details
as the gas density in the clouds and their covering factor, 
it is anyway suggestive that the X-rays emitted from the `far side' 
of the stellar wind (as seen by an observer) have a higher chance to 
suffer absorption than those emitted from the `near side'.
Therefore,  the emergent X-ray emission might be altered appreciably, 
resulting in a blue-shift of the spectral lines. 
 
The specific stellar
`cloudiness' defined above is proportional to the quantities
that determine the X-ray absorption efficiency of the cooling
zones in the clumps and the cold clouds:
namely, the geometrical extension of a cloud (the quantity $\delta$) and
the cold-gas density. The latter is proportional to the density of 
the smooth
stellar wind, that is to the ratio of the mass loss to the wind velocity.
Thus, one can use the total stellar `cloudiness' 
\footnote{{\it Total stellar `cloudiness':}
The distribution of emission measure of radiative shocks is
derived from the spectral fit of a given star. This yields the values of
$\Gamma_{sh}$ for each temperature ($T_{sh}$). Then, based
on eq.~(\ref{eq:norm}) these values are used to derive the corresponding 
specific `cloudiness' ($\delta$) for each radiative shock in the
distribution. Finally, the individual $\delta$-values are added to
give the total stellar `cloudiness' for the studied object.
}
in a given object, 
given by the sum of all the specific values, as a measure of the 
overall efficiency of the cold X-ray absorbers in the wind. 
In such a case, a correlation 
between the total 'cloudiness' and the line shift of the spectral lines 
should be expected, 
in the sense that the larger the total `cloudiness', the higher the 
amount of blue-shift of the X-ray spectrum of the star.

Figure~\ref{fig:cloudy} presents the results of our estimate of the total
stellar `cloudiness' vs. the global line shifts in the X-ray spectra
of the massive stars of our sample. 
These results indicate that the expected 
correlation does exist: winds
from stars with high values of the covering factor display a larger amount
of blue-shift in the X-ray line profiles.  
Note that for many objects the derived from the fit errors can be
appreciable.
Unfortunately, the complexity of the model fits and
the non-uniform quality of the data limit the accuracy on the derivation 
of the plotted qauntities. An additional source of uncertainty comes 
from the `cloudiness' parameter that
depends on the accuracy of the cooling curve (see eq.~\ref{eq:spec}).
However, the key point is that all the objects were analyzed in a uniform way.
Thus, future improvements in the atomic data (i.e. cooling curve) might only
shift the scale, but they will not affect the observed trend that we believe 
is rather robust.
More importantly, as can be seen in Fig.~\ref{fig:cloudy}  appreciable 
red shifts of the spectral lines are not observed.

The observed correlation between the total stellar `cloudiness' and
the spectral line shifts reveals additional details about the X-ray
emitting zone in hot massive stars. 
For example, from Fig.~\ref{fig:cloudy} and for typical mass-loss rates
($\dot{M}_6 \approx 0.1 - 1$ ) and  wind velocities ($v_{1000}
\approx 1.5 - 2.5$), in most cases the `cloudiness' parameter is 
small ($\delta \leq 1$).
In other words, dense clumps/clouds do not completely cover the
X-ray `sky' of massive stars. 
This also means that the RDI instabilities (and the MCWS as well) 
do not affect a big part of their stellar winds.

Comparing the results for single and 
multiple systems, we 
note that the latter show higher values of the `cloudiness' parameter: 
this could result from
additional sources of X-ray emission, such as 
colliding wind shocks in binary
stars (\citealt{luo_90}; \citealt{st_92}; \citealt{myas_93}).
We have to keep in mind that the
`cloudiness' parameter is also a measure of the efficiency of
the X-ray production.
Therefore, if an extra source of X-rays exists in some object, then it 
will result in an effectively larger number of shocks representing the
additional emission. 
This is the case for a binary system, where
a higher value of that parameter is expected
in comparison to a single star with the same spectral line shift. 

Finally, the fact that no appreciable redshift of the X-ray line profiles 
is observed in any object indicates that the distribution of the absorbers 
in OB stars likely has a spherical (or axial) symmetry. Future observations 
with a high spectral resolution and a good photon statistics will
give us an opportunity to further test these conclusions and 
to study the X-ray gas kinematics in greater detail.
They will also show whether the `cloudiness' parameter for
a given object changes with time which might be an important
characteristic of the mechanism that drives the wind instabilities 
responsible for the X-ray emission from massive stars.

\section{Conclusions}
\label{sec:con}
Using the {\it Chandra} public archive,
we have analyzed the X-ray spectra of a sample of 15 massive OB stars.
The basic assumption in this study is that the X-ray emission from such
objects originates in shocks which develop in the stellar wind either as
a result of radiation-driven instabilities or due to confinment of the
wind by relatively strong magnetic field. 
The main results and conclusions are as follows:

1. Based on the fast shock cooling in the winds of hot massive stars, 
we have developed a simple model
of a steady-state, plane-parallel radiative shock which has been 
incorporated in the software package for analysis of X-ray 
spectra, \XSPEC. 

2. The model is then used in a global analysis
of the X-ray emission from the sample of OB stars with high quality spectra
obtained by {\it Chandra}. We assume that the X-ray emission originates
from an ensemble of shocks that are most efficient in the acceleration
zone of the stellar wind. Far from the formation region the shock 
structures (clumps, complete or fractional shells) cool down and  
eventually become `cold' clouds in the X-ray sky of the star. In case of high
coverage,  these clouds can effectively absorb the X-rays emitted by the 
underlying shocks.
In such a case, the spectral lines in the X-ray spectrum of the massive star
will be blue-shifted. 

3. Using our model, we define a new quantity, the
stellar `cloudiness', which represents the X-ray absorption efficiency
by the clouds in the stellar wind. Our analysis
reveals that despite the large intrinsic uncertainties, there is
a correlation between the stellar `cloudiness' and the global line
blue-shifts in the observed X-ray spectra of the sample stars.

4. For each star the model gives the distribution of 
emission measure (DEM) of the radiative shocks responsible for the 
observed X-rays. 
The sample of OB stars with good spectra is
distinguished in two groups: (i) one with the DEM peaked at $\sim$0.1-0.4 keV 
and with a very low level of high energy tail above 1.0 keV;
(ii) one with the maximum of the DEM that falls at considerably higher 
temperatures. In the former case, 
the radiative-driven instability shocks are the
likely mechanism for the production of X-rays, while in the latter 
($\theta^1$ Ori C, $\tau$ Sco and $\zeta$ Oph) the X-ray emission
originates in magnetically-confined wind shocks.

5. The derivation of metal abundance indicates a subsolar metallicity
for all the stars of the sample, with iron a factor of 2-3 below the solar value.

\section{Acknowledgments}
Partial financial support from the Bulgarian Academy of Sciences - Consiglio
Nazionale delle Ricerche bilateral cooperation programme is acknowledged.
This research has made use of the NASA's Astrophysics Data System,
and the SIMBAD astronomical database
operated by CDS at Strasbourg, France.

\appendix
\section{The Model}
\label{app}
The physical background of the radiative shock model was given in 
\S~\ref{sec:mod} along with the basic features of our global model that 
assumes a distribution of radiative shocks in the stellar wind of massive 
OB stars. Some technical details about these two models are described 
in this Appendix.

\underline{{\it Shock Model:}} 
As discussed in \S~\ref{subsec:shock}, the X-ray spectrum of a parcel
of gas downstream from the shock front is given by the emission
from an optically-thin plasma in CIE. Thus, the total spectrum of a
radiative shock is an integral over a range of plasma temperatures,
$T \leq T_{sh}$, in the cooling zone (see eq.~[\ref{eq:spec}]), where
$T_{sh}$ is the immediate postshock temperature for a strong shock.
As seen from eq.~(\ref{eq:spec}), the emission measure
in the cooling zone is a power-law function of temperature, $T$, and the 
spectrum at each value of $T$ is assumed to be that given by the 
{\it vapec} model in \XSPEC. The technical realization is based on the 
\XSPEC subroutine 
{\it sumdem} with the APEC switch ON. For convenience, the numerical
integration is over common logarithms ($d \lg T$). The X-ray emission
from plasma cooler than $\sim 300,000$~K is not taken into account. 

Thus, our \XSPEC model of the X-ray emission from a radiative shock
has the following parameters: the postshock temperature $T_{sh}$,
chemical abundances, and the normalization factor $\Gamma_{sh}$. As
for all optically-thin plasma models in \XSPEC, the model normalization is
related to the emission measure of the hot plasma and in this case 
$\Gamma_{sh}$ gives the total EM in a radiative shock
(for temperatures $T \leq T_{sh}$).

\underline{{\it Global Model:}}
Our global model assumes that the X-rays from massive OB stars
originate in an ensemble of radiative shocks (\S~\ref{subsec:global}).
We use Chebyshev polinomials algorithm (\citealt{le_89})
to describe the smoothed distribution of the emission measure 
in this ensemble of shocks. Since the basic parameter for shocks is
the immediate postshock temperature, this distribution is computed
relative to $T_{sh}$.
Thus, the {\it total} EM in each shock of the ensemble is
derived from the spectral fits.
Note that the situation is similar in the \XSPEC model {\it c6pvmkl} 
(which is based on \citealt{le_89})
where the basic vectors are those giving the
X-ray emission (spectrum) from 
optically-thin plasma in CIE. 
Here, the difference is that at each temperature $T_{sh}$ we have
the X-ray spectrum from a radiative shock, as described in 
\S~\ref{subsec:shock} and above.
In the technical realization of the global shock model the total
X-ray spectrum is derived by integrating the spectra from radiative 
shocks over the range of postshock temperatures. For convenience, this 
integration is over common logarithms ($d \lg T_{sh}$).

It should be noted that there is one important qualitative difference
between the use of the global shock model or the model with distribution 
of emission measure of optically-thin plasma (e.g., {\it c6pvmkl} in
\XSPEC) for interpreting the observed X-ray spectra.
In the former case, the EM at a given temperature, $T_{sh}$, 
in the distribution is the total EM (including plasma with $T \leq
T_{sh}$)
in a radiative shock  
with velocity corresponding 
to this postshock temperature. 
In the latter case, the EM is that of an isothermal hot gas at 
that given temperature in the distribution. 
If an object is studied whose X-ray emission originates in radiative
shocks, as is the case of massive OB stars, and a model
like {\it c6pvmkl} in \XSPEC is adopted, it is not quite correct to assign a
shock velocity to each temperature value in the hot-gas distribution. 
This is because
the hot plasma in a radiative shock is not isothermal
(\S~\ref{subsec:shock}). Thus, the amount of hot gas at each temperature 
in the {\it c6pvmkl} distribution is the total contribution
from all radiative shocks having some plasma at that temperature.
This limitation is not present in our global shock model discussed 
in this work.

Additional advantage of the radiative shock model is that its
normalization factor, $\Gamma_{sh}$, which gives the total emission
measure in the shock, can be expressed through some of the basic
shock parameters (as done in deriving eq.~[\ref{eq:spec}] using
eq.~[\ref{eq:energy}]). 
In the
simplified picture of spherically-symmetric stellar winds, it is
justified to assume that the shocks move radially, and their surface
can be presented as a fraction, $\delta$, of the `local' sphere:
$A = \delta 4\pi R^2$. Thus, the normalization parameter,
$\Gamma_{sh}$,
of our \XSPEC model can be written:
\begin{eqnarray}
  \Gamma_{sh} &=& 
\frac{5A(1+x_e)n_{sh}v_{sh}k}{2\Lambda_0}
\frac{10^{-14}}{4 \pi d^2} \nonumber \\
             &=&
       2.444~\delta \left( \frac{\dot{M}_6}{v_{1000}} \right)
       \frac{\sqrt{T_{sh}}}{d^{~2}_{kpc}}
\label{eq:norm}
\end{eqnarray}
where $d_{kpc}$ is the distance to the
object in kpc, and all other quantities
are defined in \S~\ref{sec:mod}.
The quantity ($10^{-14}/4 \pi d^2$) comes from the \XSPEC 
normalization factor of the flux for optically-thin plasma models.
Note that $\Gamma_{sh}$ does not depend on the shock radius 
since in the case under consideration the gas density, $n_{sh}$, 
goes as $R^{-2}$.
Again, the relative number density of
hydrogen is assumed $x_H = 0.9$ and $x_e = 1.1$ for a fully
ionized plasma. For simplicity, in deriving eq.~(\ref{eq:norm})
we replaced the local wind velocity with the terminal speed
which means that the effective surface of the shock
(presented by $\delta$) has the `maximum' value. We note that the
quantity $\delta$
may have values larger than unity since it represents the total area
of the shocks with a particular velocity within the entire ensemble
of shocks.
Thus, once the value of the flux normalization parameter,
$\Gamma_{sh}$, is derived from a spectral fit, it can be further used
to gain insight into the physical picture,
provided some of the stellar parameters 
are already known from other studies.

\end{document}